\documentclass[%
preprint,
nofootinbib,
amsmath,amssymb,
aps,
prd,
]{revtex4-1}
\usepackage{setspace}
\usepackage[caption=false]{subfig}
\usepackage{graphicx}
\usepackage{dcolumn}
\usepackage{bm}
\usepackage{mathtools}
\DeclarePairedDelimiter{\ceil}{\lceil}{\rceil}

\usepackage{color}

\usepackage[unicode=true,bookmarks=true,
bookmarksnumbered=false,bookmarksopen=false,
breaklinks=false,
pdfborder={0 0 1},
backref=false,colorlinks=true]{hyperref}
\hypersetup{pdftitle={Consistency Conditions for an AdS/MERA Correspondence}, pdfauthor={Ning Bao,  ChunJun Cao, Sean M. Carroll, Aidan Chatwin-Davies, Nicholas Hunter-Jones, Jason Pollack, and Grant N. Remmen}, citecolor=blue,linkcolor=blue,urlcolor=blue,citecolor=blue}

\newcommand{\pr}{\prime}

\newcommand{\bra}[1]{\langle #1|}
\newcommand{\ket}[1]{|#1\rangle}

\newcommand{\be}{\begin{equation}}
\newcommand{\ee}{\end{equation}}
\newcommand{\bea}{\begin{eqnarray}}
\newcommand{\eea}{\end{eqnarray}}
\newcommand{\Eq}[1]{Eq.~(\ref{#1})}
\newcommand{\Eqs}[2]{Eqs.~(\ref{#1}) and (\ref{#2})}
\newcommand{\Sec}[1]{Sec.~\ref{#1}}

\newcommand{\Fig}[1]{Fig.~\ref{#1}}
\newcommand{\App}[1]{App.~\ref{#1}}

\newcommand{\Ref}[1]{Ref.~\cite{#1}}
\newcommand{\Refs}[1]{Refs.~\cite{#1}}
\begin{document}

\preprint{CALT-TH-2015-015}

\title{Consistency Conditions for an AdS/MERA Correspondence}

\author{Ning Bao,  ChunJun Cao, Sean M. Carroll, Aidan Chatwin-Davies, Nicholas Hunter-Jones, Jason Pollack, and Grant N. Remmen}
\thanks{ \vspace*{-3mm}
\href{mailto:ningbao@theory.caltech.edu}{\tt ningbao@theory.caltech.edu},
\href{mailto:seancarroll@gmail.com}{\tt seancarroll@gmail.com},
\href{mailto:cjcao@caltech.edu}{\tt cjcao@caltech.edu},\\ \vspace*{-3mm}
\href{mailto:achatwin@caltech.edu}{\tt achatwin@caltech.edu},
\href{mailto:nickrhj@theory.caltech.edu}{\tt nickrhj@theory.caltech.edu},
\href{mailto:jpollack@caltech.edu}{\tt jpollack@caltech.edu},\\ \vspace*{-3mm}
\href{mailto:gremmen@theory.caltech.edu}{\tt gremmen@theory.caltech.edu}
}

\affiliation{
Walter Burke Institute for Theoretical Physics,\\ California Institute of Technology,
Pasadena, CA 91125
}

\begin{abstract}

The Multi-scale Entanglement Renormalization Ansatz (MERA) is a tensor network that provides an efficient way of variationally estimating the ground state of a critical quantum system. The network geometry resembles a discretization of spatial slices of an AdS spacetime and ``geodesics'' in the MERA reproduce the Ryu--Takayanagi formula for the entanglement entropy of a boundary region in terms of bulk properties. It has therefore been suggested that there could be an AdS/MERA correspondence, relating states in the Hilbert space of the boundary quantum system to ones defined on the bulk lattice. Here we investigate this proposal and derive necessary conditions for it to apply, using geometric features and entropy inequalities that we expect to hold in the bulk. We show that, perhaps unsurprisingly, the MERA lattice can only describe physics on length scales larger than the AdS radius.  Further, using the covariant entropy bound in the bulk, we show that there are no conventional MERA parameters that completely reproduce bulk physics even on super-AdS scales. We suggest modifications or generalizations of this kind of tensor network that may be able to provide a more robust correspondence.

\end{abstract}


\maketitle

\tableofcontents

\section{Introduction}

The idea that spacetime might emerge from more fundamental degrees of freedom has long fascinated physicists. The holographic principle suggests that a $(D+1)$-dimensional spacetime might emerge from degrees of freedom in a $D$-dimensional theory without gravity \cite{holo1,holo2}. While a completely general implementation of this idea is still lacking, the AdS/CFT correspondence provides a specific example in which to probe the holographic emergence of spacetime. AdS/CFT is a conjectured correspondence between $D$-dimensional conformal field theories (CFTs) in Minkowski space and $(D+1)$-dimensional asymptotically anti-de Sitter (AdS) spacetimes \cite{MaldacenaAdSCFT,WittenAdSCFT,MAGOO}.
An intriguing aspect of this duality is the Ryu--Takayanagi formula \cite{RyuTakayanagi,LewkMald13}, according to which the entanglement entropy of a region $B$ on the boundary is proportional to the area of a codimension-two extremal surface $\tilde B$ embedded in the bulk curved spacetime whose boundary is $B$:
\be 
S(B) = \frac{\mathrm{area}(\tilde{B})}{4G} + \text{corrections}.
\ee 
In other words, given a CFT state, one may think of bulk distance and geometry (at least near the boundary) as being charted out by the entanglement properties of the CFT state.

A central question in this picture of spacetime emerging from entanglement is: What is the precise relationship between bulk degrees of freedom and boundary degrees of freedom? Expressed in a different way, what is the full map between states and operators in the boundary Hilbert space and those in the bulk? While investigations of AdS/CFT have thrown a great deal of light on this question, explicit simple models are still very helpful for studying it in more detail.

Meanwhile, from a very different perspective, tensor networks have arisen as a useful way to calculate quantum states in strongly-interacting many-body systems \cite{Orus2014}. One significant example is the Multi-scale Entanglement Renormalization Ansatz (MERA)~\cite{Vidal}, which is relevant for critical (gapless) systems, \emph{i.e.}, CFTs. Starting from a simple state in a low-dimensional Hilbert space, acting repeatedly with fixed tensors living on a network lattice produces an entangled wave function for the quantum system of interest; varying with respect to the tensor parameters efficiently computes the system's ground state. 

Working ``backwards'' in the MERA, starting with the ground state and gradually removing entanglement, produces a set of consecutively renormalized quantum states. This process reveals a renormalization direction along the graph, which may be thought of as an emergent radial direction of space. As pointed out by Swingle \cite{Swingle2012}, the MERA graph can serve as a lattice discretization of spatial slices of AdS. Furthermore, one can use the MERA to calculate the entanglement entropy of regions of the original (boundary) critical system; this calculation amounts to tracing over bonds in the tensor network that cross the causal cone of the boundary region. The causal cone is a sort of extremal surface for the MERA, motivating comparison to the Ryu--Takayanagi formula.

It is therefore natural to conjecture that the MERA provides a concrete implementation of the emergence of spacetime, in the form of a correspondence between boundary and bulk regions reminiscent of AdS/CFT \cite{Swingle2012}. Such an AdS/MERA correspondence would be extremely useful, since the basic building blocks of the MERA are discrete quantum degrees of freedom from which quantities of physical interest may be directly calculated. Some specific ideas along these lines have recently been investigated \cite{Qi2013,HQECCs,MERAintgeo,Beny2013}.

In this paper, we take a step back and investigate what it would mean for such a correspondence to exist and the constraints it must satisfy in order to recover properties we expect of physics in a bulk emergent spacetime. After reviewing the MERA itself and possible construals of the AdS/MERA correspondence in the next section, in \Sec{sec:geodesics} we then derive relationships between the MERA lattice and the geometry of AdS. We find that the MERA is unable to describe physics on scales shorter than the AdS radius. In \Sec{sec:entanglement} we explore constraints from calculating the entanglement entropy of  regions on the boundary, in which we are able to relate MERA parameters to the central charge of the CFT. Finally, in \Sec{sec:BH} we apply the covariant entropy (Bousso) bound to regions of the bulk lattice. In the most na\"ive version of the AdS/MERA correspondence, we find that no combination of parameters is consistent with this bound, but we suggest that generalizations of the tensor network may be able to provide a useful correspondence.

\section{\label{sec:adsmera}AdS/MERA}

Let us begin by recalling the definition and construction of the MERA. We will then introduce the AdS/MERA correspondence and discuss the motivation for and consequences of this proposal.

\subsection{Review of the MERA}

The MERA is a particular type of tensor network that provides a computationally efficient way of finding the ground states of critical quantum many-body systems, \emph{i.e.} CFTs, in $D$ dimensions. (For a recent review of tensor networks in general, see \Ref{Orus2014}. Detailed analyses of the MERA are given in \cite{Vidal,MERAAlg,MERACFT} and references therein.) In this work, we restrict our attention to the case $D=1+1$.

\begin{figure}
\vspace{-1.3cm}
\subfloat[]{
  \includegraphics[width=0.75\textwidth]{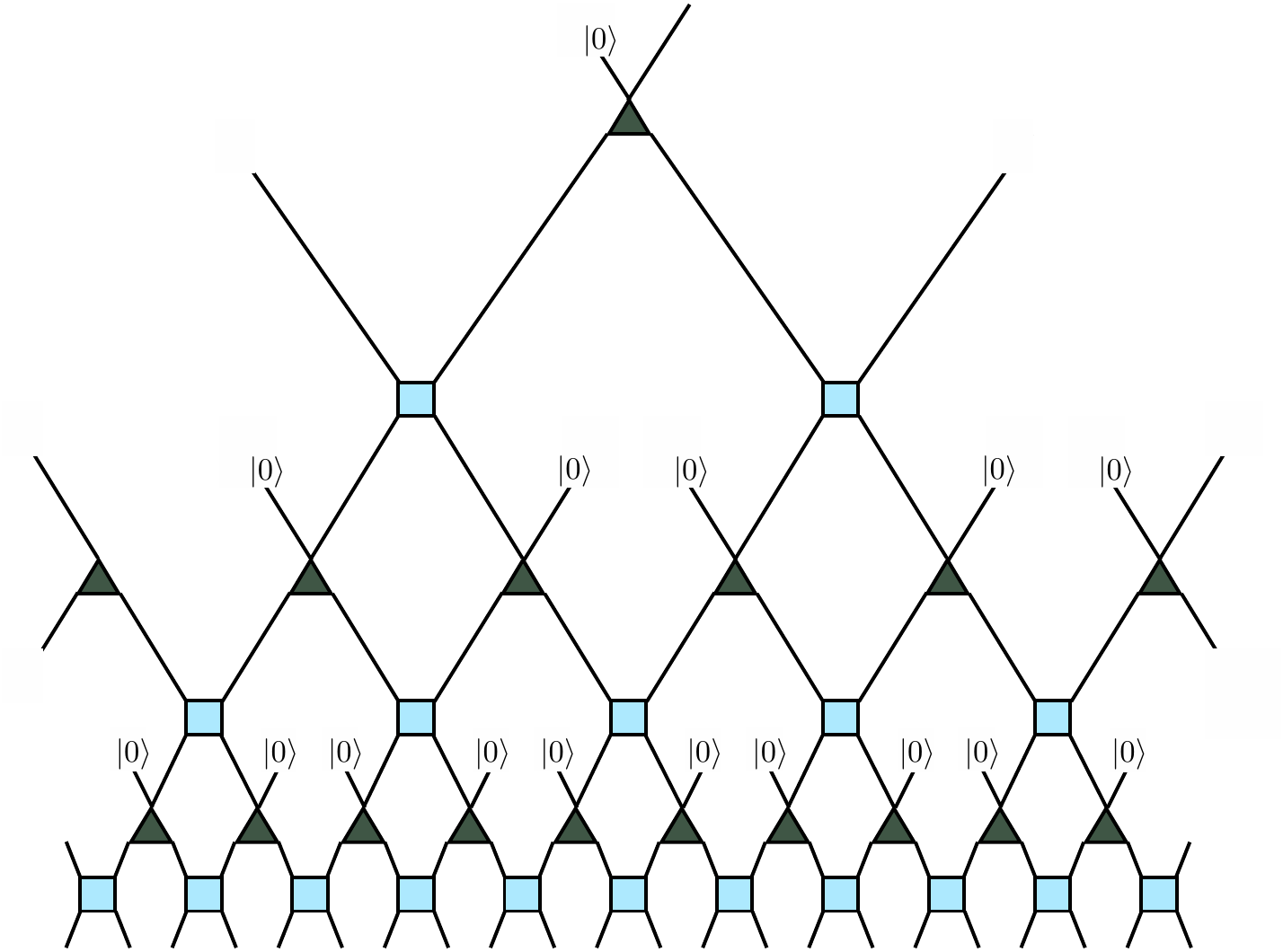}
  \label{fig:full2to1mera}
}
\begin{minipage}[b][11cm][t]{.25\textwidth}
  \vspace*{\fill}
  \centering
  \subfloat[]{\includegraphics[width=0.5\textwidth]{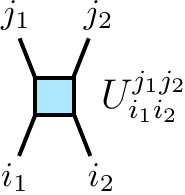}
  \label{fig:disentangler}}
  \par\vfill
  \subfloat[]{\includegraphics[width=0.5\textwidth]{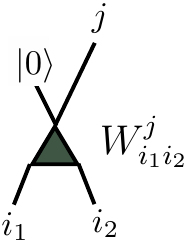}
  \label{fig:isometry}}
\end{minipage}

\caption{(a) Basic construction of a $k=2$ MERA (2 sites renormalized to 1). (b) The squares represent disentanglers: unitary maps that, from the moving-upward perspective, remove entanglement between two adjacent sites. (c) The triangles represent isometries: linear maps that, again from the moving-upward perspective, coarse-grain two sites into one. Moving downward, we may think of isometries as unitary operators that, in the MERA, map a state in $V \otimes \ket{0}$ into $V \otimes V$. The $i$ and $j$ labels in (b) and (c) represent the tensor indices of the disentangler and isometry. }\label{fig:full2to1meraSUPER}
\end{figure}

The MERA tensor network is shown in \Fig{fig:full2to1meraSUPER}. The quantum system being modeled by the MERA lives at the bottom of the diagram, henceforth ``the boundary'' in anticipation of the AdS/MERA connection to be explored later. We can think of the tensor network as a quantum circuit that either runs from the top down, starting with a simple input state and constructing the boundary state, or from the bottom up, renormalizing a boundary state via coarse-graining. One defining parameter of the MERA is the rescaling factor $k$, defining the number of sites in a block to be coarse-grained; in \Fig{fig:full2to1meraSUPER} we have portrayed the case $k=2$. 
The squares and triangles are the tensors: multilinear maps between direct products of vector spaces. Each line represents an index $i$ of the corresponding tensor, ranging over values from $1$ to the ``bond dimension'' $\chi$. 
The boundary Hilbert space $\mathcal{H}_\text{boundary} = V^{\otimes {\mathcal{N}_\text{boundary}}}$ 
is given by a tensor product of $\mathcal{N}_\text{boundary}$ individual spaces $V$, each of dimension $\chi$. 
(In principle the dimension of the factors in the boundary could be different from the bond dimension of the MERA, and indeed the bond dimensions could vary over the different tensors. We will assume these are all equal.) 

As its name promises, the MERA serves to renormalize the initial boundary state via coarse-graining. If we were to implement the MERA for only a few levels, we would end up with a quantum state in a smaller Hilbert space (defined on a fixed level of the tensor network), retaining some features of the original state but with some of the entanglement removed. However, we can also run the MERA backwards, to obtain a boundary state from a simple initial input. By varying the parameters in the individual tensors, we can look for an approximation of the ground state of the CFT on the boundary. Numerical evidence indicates that this process provides a computationally efficient method of constructing such ground states \cite{MERACFT,MERACFTNonlocal}. 

The tensors, or gates, of the MERA come in two types. The first type are the disentanglers, represented by squares in \Fig{fig:full2to1meraSUPER}. These are unitary maps $U \!: V_{} \otimes V_{} \rightarrow V_{} \otimes V_{}$, as in \Fig{fig:disentangler}. The name comes from thinking of moving upward through the network, in the direction of coarse-graining, where the disentanglers serve to remove local entanglement; as we move downward, of course, they take product states and entangle them.
The second type of tensors are the isometries, represented by triangles. 
From the moving-downward perspective these are linear maps $W\!: V \rightarrow V_{} \otimes V_{}$; moving upward, they implement the coarse-graining, see \Fig{fig:isometry}. The isometries  are subject to the further requirement that $W^\dagger W = I_{V}$, where $I_{V}$ is the identity map on $V$, and $WW^\dagger = P_{A_{}}$, where $P_{A_{}}$ is a projector onto some subspace ${A_{}} \subset V_{} \otimes V_{}$. 
From the top-down perspective, we can also think of the isometries as bijective unitary operators $W_U\!:V_{} \otimes V_{} \rightarrow V_{} \otimes V_{}$, for which a fixed ``ancilla'' state (typically the ground state $|0\rangle$) is inserted in one of the input factors, as shown in \Fig{fig:isometry}.
More generally, isometries could map $q<k$ sites onto $k$ sites, $W\!:  V^{\otimes q} \rightarrow V^{\otimes k}$. 

The MERA is not the simplest tensor network which implements coarse-graining. For instance, the tree tensor network \cite{TTN} (also considered in a holographic context in \Ref{Qi2013}), similar to MERA but without any disentanglers, also implements coarse-graining. However, tensor networks without disentanglers fail to capture the physics of systems without exponentially-decaying correlations, and consequently  cannot reproduce a CFT ground state.

An example that invites analysis with a MERA is the transverse-field Ising model \cite{TransIsing}. In $1+1$ dimensions, the model describes a chain of spins with nearest-neighbor interactions subject to a transverse magnetic field. Its Hamiltonian is
\be 
\hat H = -J \sum_{i} \hat \sigma_i^z \hat \sigma_{i+1}^z - h \sum_{i} \hat \sigma_i^x \, ,
\ee 
where $\hat \sigma_i^z$ and $\hat \sigma_i^x$ are Pauli operators and where $J$ and $h$ set the strength of the nearest-neighbor interactions and the magnetic field, respectively. Notably, the system achieves criticality at $J = h$, where a quantum phase transition occurs between ordered ($J>h$) and disordered ($J<h$) phases. In this example, the open legs at the bottom of the MERA describe the state of the one-dimensional lattice of spins. A single application of disentanglers and isometries can be thought of as a true real-space renormalization, producing a lattice of spins that is less dense than the preceding lattice by a factor of $q/k$. 

In general, much information is required to describe an arbitrary MERA. In principle, the Hilbert spaces, the disentanglers, and the isometries could all be different. Also, for $k>2$, there is no canonical way of laying out the disentanglers and isometries; the circuit itself must be specified. We will restrict ourselves to the case $q=1$, so that isometries have $1$ upward-going leg and $k$ downward-going legs. Further, without loss of generality, we take the same vector spaces, disentanglers, and isometries everywhere in the MERA, a simplification that is enforced by the symmetries of the boundary ground state. These symmetries --- namely, translation- and scale-invariance --- dictate that the MERA parameters and structure be homogeneous across the whole tensor network.

\begin{figure}[ht]
\centering
\subfloat[]{
\includegraphics[scale=0.45,width=0.42\textwidth]{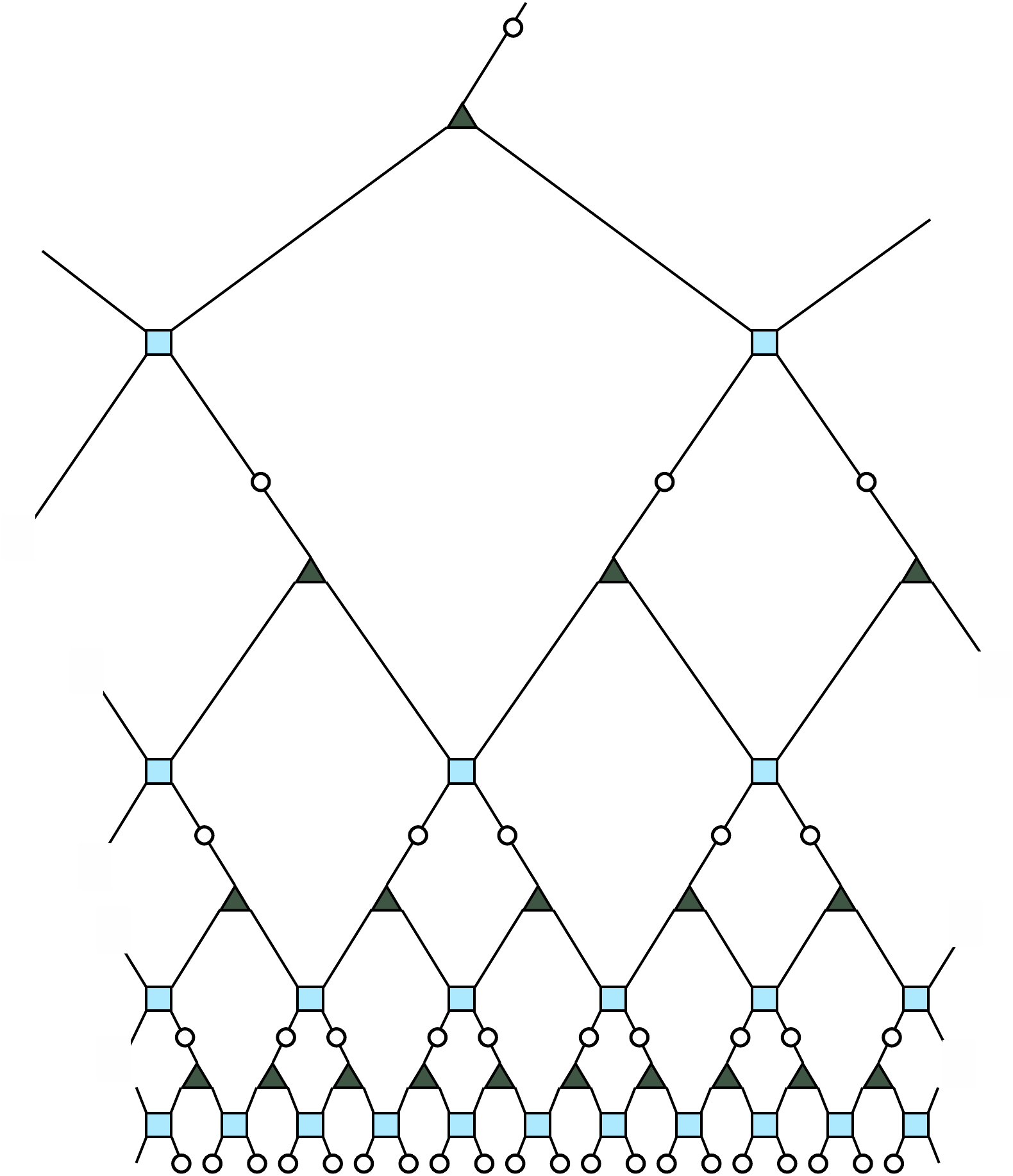}
}
\subfloat[]{
\includegraphics[scale=0.7,width=0.5\textwidth]{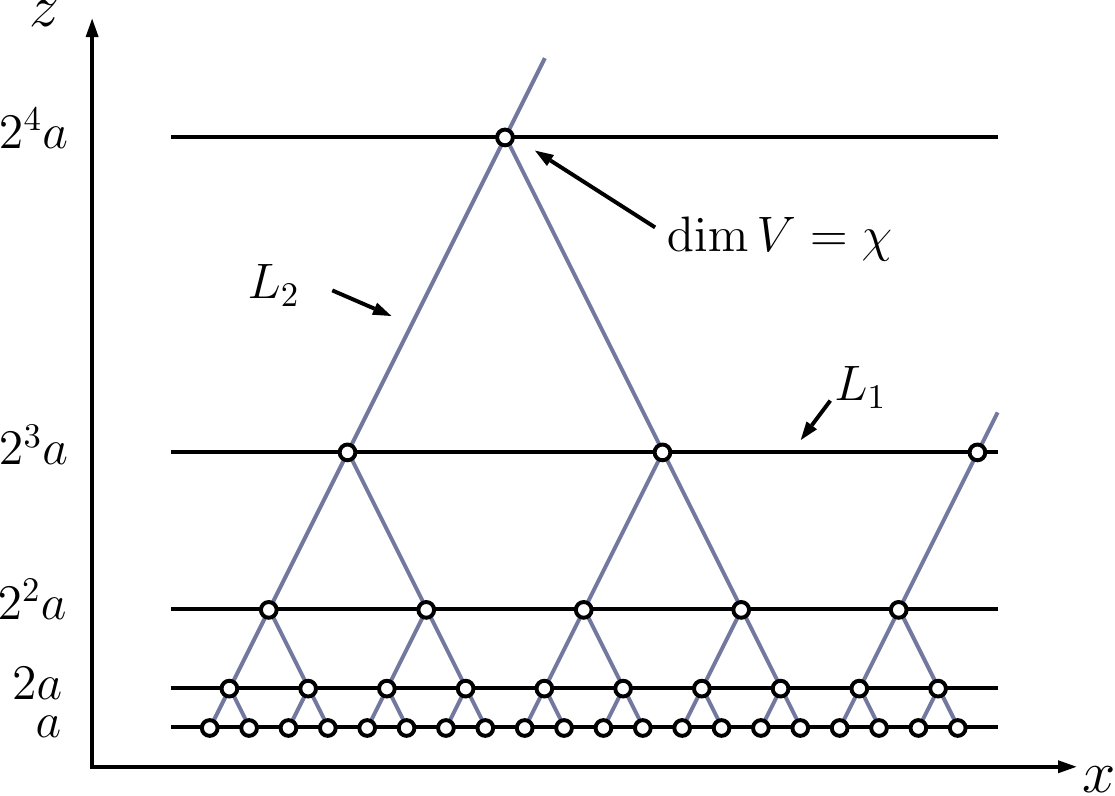}
}
\caption{(a) A $k=2$ MERA, and (b) the same MERA with its disentanglers and isometries suppressed. The horizontal lines in the graph on the right indicate lattice connectivity at different renormalization depths, and the vertical lines indicate which sites at different depths are related via coarse-graining due to the isometries. Each site, represented by a circle, is associated with a Hilbert space $V$ with bond dimension $\chi$. In the simplest case, a copy of the same Hilbert space is located at each site. When assigning a metric to the graph on the right, translation and scale invariance dictate that there are only two possible length scales: a horizontal proper length $L_1$ and a vertical proper length $L_2$. }
\label{fig:2to1mera}
\end{figure}

For geometric considerations, it is useful to abstract away all of the information about unitary operators and to draw a MERA as a graph as shown in \Fig{fig:2to1mera}. In such a graph, we only indicate the connectivity of sites at any given level of coarse-graining as well as the connectivity of sites under renormalization group flow.

\subsection{An AdS/MERA Correspondence? \label{sec:AdSMERAcor}}

The possibility of a correspondence between AdS and the MERA was first proposed by Swingle in \Ref{Swingle2012}, where it was noted that the MERA seems to capture certain key geometric features of AdS. At the most basic level, when viewed as a graph with legs of fixed length, a MERA may be thought of as a discretization of the hyperbolic plane, which is a spatial slice of AdS$_3$. In this discretization, the base of the MERA tree lies on the boundary of the AdS slice and the MERA lattice sites fill out the bulk of the slice \cite{Swingle2012, MaldHart}.

Interestingly, the structure of a MERA is such that it seems to go beyond a simple discretization of the hyperbolic plane. Certain discrete paths in the MERA naturally reproduce geodesics of the hyperbolic plane \cite{Swingle2012,Swingle2012a}. Moreover, this phenomenon makes it possible to understand the computation of CFT entanglement entropy using a MERA as a discrete realization of the Ryu--Takayanagi formula \cite{Evenbly2014}. These and other examples \cite{Swingle2012,Swingle2012a} seem to suggest that a MERA may in fact be elucidating the structural relationship between physics on the boundary of AdS and its bulk.

In this work we take the term ``AdS/MERA correspondence" to mean more than simply a matching of graph geometry and continuous geometry. In the spirit of the AdS/CFT correspondence, we suppose that (at least some aspects of) both boundary and bulk physics are described by appropriate Hilbert spaces $\mathcal{H}_\text{boundary}$ and $\mathcal{H}_\text{bulk}$ respectively, which must have equal dimensions. A full AdS/MERA correspondence would then be a specification of these Hilbert spaces, as well as a prescription which makes use of the MERA to holographically map states and operators in $\mathcal{H}_\text{boundary}$ to corresponding states and operators in $\mathcal{H}_\text{bulk}$ and vice-versa. 
To preserve locality in the bulk and the symmetries of AdS, it is natural to identify $\mathcal{H}_\text{bulk}$ with the tensor product of individual spaces $V_\text{bulk}$, each located at one site of the MERA.
If it exists, this correspondence provides a formulation of bulk calculations in terms of the MERA. An AdS/MERA correspondence should allow us to, for example, calculate bulk correlation functions, or bulk entanglement entropies using tools from or the structure of the MERA.

There is one straightforward way to construct such a map $\mathcal{H}_\text{boundary} \leftrightarrow \mathcal{H}_\text{bulk}$. We have noted that the isometries $W\! : V \rightarrow V\otimes V$ can be thought of as unitaries $W_U\! :V\otimes V \rightarrow V\otimes V$ by imagining that a fixed ancillary state $|0\rangle$ is inserted in the first factor; for a $k$-to-one MERA, one would insert $k-1$ copies of the $|0\rangle$ ancilla at each site to unitarize the isometries. From that perspective, running upwards in the tensor network provides a map from the MERA ground state on the boundary to a state $\ket{0}^{\otimes(k-1)\mathcal{N}_\text{bulk}} \in V^{\otimes(k-1)\mathcal{N}_\text{bulk}}$, where at each isometry there is a copy of $V^{\otimes(k-1)}$ and $\mathcal{N}_\text{bulk}$ denotes the number of bulk lattice sites, excluding the boundary layer. As we ultimately show in \Sec{sec:BH}, one has $\mathcal{N}_\text{boundary} = (k-1)\mathcal{N}_\text{bulk}$. We can then identify $\mathcal{H}_\text{boundary} = \mathcal{H}_\text{bulk} = V^{\otimes \mathcal{N}_\text{boundary}}$ and think of the tensor network as a quantum circuit providing a map between arbitrary states $\mathcal{H}_\text{boundary}\rightarrow \mathcal{H}_\text{bulk}$. In this construction, the MERA ground state on the boundary gets mapped to the factorized bulk state $\ket{0}^{\otimes (k-1)\mathcal{N}_\text{bulk}}$, but other boundary states will in general produce entangled states in the bulk (keeping the tensors themselves fixed).

Something very much like this construction was proposed by Qi \cite{Qi2013}, under the name ``Exact Holographic Mapping" (EHM). That work examined a tensor network that was not quite a MERA, as no disentanglers were included, only isometries. As a result, while there is a map $\mathcal{H}_\text{boundary}\rightarrow \mathcal{H}_\text{bulk}$, the boundary state constructed by the tensor network does not have the entanglement structure of a CFT ground state. In particular, it does not seem to reproduce the Ryu--Takayanagi formula in a robust way. Alternatively, we can depart from Qi by keeping a true MERA with the disentanglers left in, in which case the bulk state constructed by the quantum circuit has no entanglement: it is a completely factorized product of the ancilla states. Such a state doesn't precisely match our expectation for what  a bulk ground state should look like, since there should be at least some entanglement between nearby regions of space.

Therefore, while it is relatively simple to imagine constructing a bulk Hilbert space and a map between it and the boundary Hilbert space, it is not straightforward to construct such a map that has all of the properties we desire. It might very well be possible to find such a construction, either by starting with a slightly different boundary state, or by adding some additional structure to the MERA. 

For the purposes of this paper we will be noncommittal. That is, we will imagine that there is a bulk Hilbert space constructed as the tensor product of smaller spaces at each MERA site, and that there exists a map $\mathcal{H}_\text{boundary}\rightarrow \mathcal{H}_\text{bulk}$ that can be constructed from the MERA, but we will not specify precisely what that map might be. We will see that we are able to derive bounds simply from the requirements that the hypothetical correspondence should allow us to recover the properties we expect of bulk physics, including the background AdS geometry and features of semiclassical quantum gravity such as the Bousso bound on bulk entropy.

\section{\label{sec:geodesics}MERA and Geometry}

If a MERA is a truly geometrical object that describes a slice of AdS, then the graph geometry of a MERA should give the same answers to geometric questions as the continuous geometry of a slice of AdS. Here, we reconsider the observation by Swingle \cite{Swingle2012,Swingle2012a} that certain trajectories on the MERA coincide with trajectories in AdS and we investigate the constraints that this correspondence places on the graph metric of the MERA. We find that a MERA necessarily describes geometry on super-AdS length scales, moreover, there is no redefinition of the MERA coordinates that results in the proper distance between MERA sites mapping to any sub-AdS length scale.  

\subsection{Consistency conditions from matching trajectories}

In order to speak of graph geometry, one must put a metric on the MERA graph, \emph{i.e.}, one must assign a proper length to each bond in the graph of \Fig{fig:2to1mera}. Presumably, the metric should originate from correlations between the sites in the MERA. In the absence of an explicit identification of the origin of the graph metric, however, at least in the case of a MERA describing the ground state of a CFT, it is sensible to identify two length scales. Explicitly, we must assign a proper length $L_1$ to horizontal bonds and a proper length $L_2$ to vertical bonds. Indeed, translational and conformal invariance guarantee that these are the only two length scales in any graph metric one can assign to a MERA for which an AdS/MERA correspondence exists. In particular, the ground state of a CFT is translation invariant, so each horizontal bond in the finest (UV-most) lattice should have the same proper length so as to respect this symmetry. Self-similarity at all scales then requires that any horizontal bond at any level of renormalization have this same proper length. There is no \emph{a priori} reason why the vertical bonds should share the proper length of the horizontal bonds and indeed we will see that their proper length will be different. However, again by self-similarity and translation invariance, all vertical bonds must be assigned the same proper length.

The observation in \Ref{Swingle2012} that certain paths in the MERA graph coincide with corresponding paths in slices of AdS is what established the possibility of an AdS/MERA correspondence. Here we will carefully examine these paths and determine what constraints the requirements that they match place on MERA parameters, \emph{i.e.}, on the bond lengths $L_1$ and $L_2$ and on the rescaling factor $k$.

\begin{figure}[ht]
\begin{center}
\includegraphics[scale=0.75]{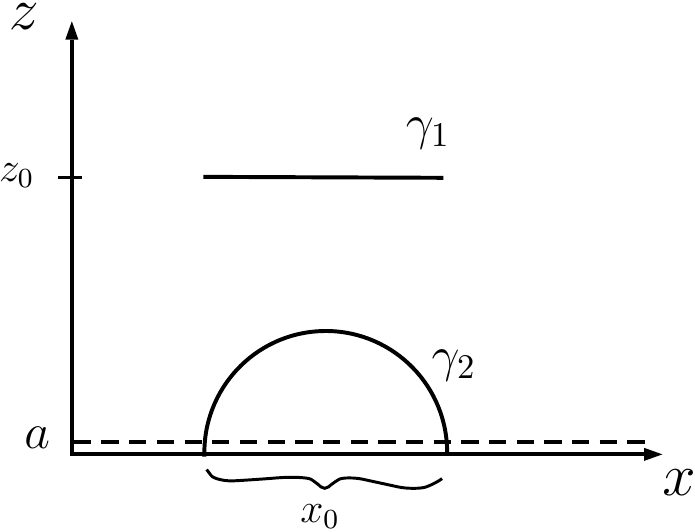}
\caption{A horizontal line ($\gamma_1$) and a geodesic ($\gamma_2$) in a spatial slice of $\mathrm{AdS}_3$. }
\label{fig:AdSlines}
\end{center}
\end{figure}

Consider a constant-time slice of $\mathrm{AdS}_3$ with the following metric:
\be  \label{eq:AdSmetric}
\mathrm{d}s^2 = \frac{L^2}{z^2}(\mathrm{d}z^2 + \mathrm{d}x^2).
\ee 
We will compare the proper lengths of straight horizontal lines and geodesics in the AdS slice to the proper lengths of the corresponding paths in the MERA graph. In the AdS slice, let $\gamma_1$ be a straight horizontal line ($\mathrm{d}z = 0$) sitting at $z = z_0$ with coordinate length $x_0$. Let $\gamma_2$ be a geodesic whose endpoints lie near the boundary $z=0$ and are separated by a coordinate distance $x_0$ at the boundary. In this choice of coordinates, such a geodesic looks like a semicircle (see \Fig{fig:AdSlines}). It is a straightforward computation to show that the proper lengths of these curves are
\be 
|\gamma_1|_\mathrm{AdS} = \frac{L}{z_0}x_0 \qquad \text{and} \qquad |\gamma_2|_\mathrm{AdS} = 2L\ln\left(\frac{x_0}{a}\right) \, .
\label{eq:AdSlengths}
\ee 
Note that there is a UV cutoff at $z = a \ll x_0$ and that we have neglected terms of order $a/x_0$.

We fix $L_1$ and $L_2$ by comparing $\gamma_1$ and $\gamma_2$ to horizontal lines and ``geodesics" in the MERA, respectively. Consider two sites in a horizontal lattice at depth $m$ (\emph{i.e.}, $m$ renormalizations of the UV-most lattice) and separated by a coordinate distance $x_0$ in the coordinate system shown in \Fig{fig:2to1mera}. By fiat, this lattice sits at $z_0 = k^m a$. The number of bonds between the two sites at depth $m$ is $x_0 / (k^m a)$ (see \Fig{fig:2to1mera} for the case $k=2$). It follows that the proper length of the line connecting the two points is just
\be 
\begin{aligned}
|\gamma_1|_{\mathrm{MERA}} &= L_1 \cdot \text{(number of bonds between endpoints)} \\
&= L_1 \left. \frac{x_0}{z_0} \right|_{z_0 = k^m a}
\end{aligned}
\label{eq:length1}
\ee 
To have $|\gamma_1|_\mathrm{AdS} = |\gamma_1|_\mathrm{MERA}$, we should therefore set $L_1 = L$.

Similarly, consider two lattice sites on the UV-most lattice separated by a coordinate distance $x_0$. If we assume that $x_0 \gg a$, then the shortest path (geodesic) in the MERA connecting the two lattice sites is the path that goes up in the renormalization direction and then back down again. The two sites are separated by $x_0/a$ bonds on the UV-most lattice, so $\log_k(x_0/a)$ renormalization steps are needed to make the sites either adjacent or superimposed. This means that the geodesic that connects the endpoints is made up of $2 \log_k(x_0/a)$ bonds (as we have to go up and then back down again, giving the factor of 2). It follows that the proper length of the geodesic is
\be 
\begin{aligned}
|\gamma_2|_\mathrm{MERA} &= L_2 \cdot \text{(number of bonds in the geodesic)} \\
&= 2L_2 \log_k\left(\frac{x_0}{a}\right)
\end{aligned}
\ee
To have $|\gamma_2|_\mathrm{AdS} = |\gamma_2|_\mathrm{MERA}$, we should therefore set $L_2 = L \ln \, k$.

\subsection{Limits on sub-AdS scale physics}

One aspect of the matching of geodesics that is immediately apparent is that the MERA scales $L_1$ and $L_2$ that parametrize the proper distance between lattice sites are of order the AdS scale $L$ or larger, as was also noted in Refs.~\cite{Swingle2012,MaldHart}. This runs counter to the typical expectation that, in a discretization of spacetime, one expects the granularity to be apparent on the UV, rather than the IR, scale. That is, sub-AdS scale locality is not manifested in the MERA construction and must be encoded within each tensor factor \cite{Swingle2012a}.

One could try to evade this difficulty by attempting to redefine the MERA coordinates $(x,z)^{\rm MERA}$ (those of \Fig{fig:2to1mera}) as functions of the AdS coordinates $(x,z)^{\rm AdS}$ (those of \Fig{fig:AdSlines}) and taking a continuum limit; above, we assumed that the two sets of coordinates were simply identified. That is, suppose $x^{\rm MERA} = f(x^{\rm AdS})$ and $z^{\rm MERA} = g(z^{\rm AdS})$. (For example, one could consider $f(x) = \varepsilon x$ for small $\varepsilon$ and imagine taking the continuum limit, with the aim of making $L_1$ much smaller than the AdS scale.) If $a$ is still the UV cutoff on the AdS side, then in the MERA we have $f(a)$ as the UV-most lattice spacing and $g(a)$ as the UV cutoff in the holographic direction. Consider the computation of $|\gamma_1|$. From the AdS side, we have $|\gamma_1|_{\rm AdS} = Lx_0^{\rm AdS}/z_0^{\rm AdS}$. On the MERA side, the number of sites spanned by $x_0^{\rm MERA} = f(x_0^{\rm AdS})$ is $x_0^{\rm MERA}/k^m f(a)$, while the holographic coordinate is $z_0^{\rm MERA} = k^m g(a)$. Hence,
\be 
|\gamma_1|_{\rm MERA} = L_1 \frac{f(x_0^{\rm AdS})}{f(a)}\frac{g(a)}{g(z_0^{\rm AdS})}.\label{eq:gamma1general}
\ee 
Equating $|\gamma_1|_{\rm AdS} = |\gamma_1|_{\rm MERA} \equiv |\gamma_1|$, we have
\be 
g(z_0^{\rm AdS}) \frac{\partial}{\partial x_0^{\rm AdS}}|\gamma_1| = L_1 \frac{f^\prime (x_0^{\rm AdS})}{f(a)}g(a) = L \frac{g(z_0^{\rm AdS})}{z_0^{\rm AdS}}.
\ee 
Since the right side of the first equality only depends on $x_0^{\rm AdS}$ and the second equality only depends on $z_0^{\rm AdS}$, but we can vary both parameters independently, both expressions must be independent of both AdS coordinates. Hence, we must have $f(x) = \varepsilon_x x$ and $g(z) = \varepsilon_z z$ for some constants $\varepsilon_x$ and $\varepsilon_z$. Plugging everything back into \Eq{eq:gamma1general} and comparing with $|\gamma_1|_\mathrm{AdS}$, we again find that $L_1 = L$, so no continuum limit is possible. Similarly, in computing $|\gamma_2|$, we note that the number of bonds between the endpoints on the UV-most lattice level is $x_0^{\rm MERA}/f(a)$, so the geodesic connecting the endpoints has $2 \log_k(x_0^{\rm MERA}/\varepsilon_x a)$ bonds. On the other hand, we have $|\gamma_2|_\mathrm{AdS} = 2L\ln(x_0^{\rm AdS}/a) = 2L\ln(x_0^{\rm MERA}/\varepsilon_x a)$. That is, in equating $|\gamma_2|_\mathrm{AdS}$ and $|\gamma_2|_\mathrm{MERA}$, we must again set $L_2 = L \ln k$. We thus also find that no continuum limit is possible in the holographic direction. That is, we have shown that there is a constant normalization freedom in the definition of each of the coordinate distances on the AdS and MERA sides of any AdS/MERA duality, but such a coordinate ambiguity is unphysical and does not allow one to take a continuum limit. One still finds that the physical MERA parameters $L_1$ and $L_2$ are AdS scale. This means that there truly is no sense in which a discrete MERA can directly describe sub-AdS scale physics without the addition of supplemental structure to replace the individual tensors. This fact limits the ability of the MERA to be a complete description of the gravity theory without such additional structure. It might be the case that one needs a field theoretic generalization of the MERA, such as continuous MERA (cMERA) \cite{cMERA,cMERART,cMERAFiniteT} or some local expansion of the individual tensors into discrete tensor networks with a different graph structure to describe sub-AdS physics, but such a significant generalization of the tensor network is beyond the scope of this work and in any case would no longer correspond to a MERA proper.

\section{\label{sec:entanglement}Constraints from Boundary Entanglement Entropy}

Because the MERA can efficiently describe critical systems on a lattice, quantities computed in the MERA on scales much larger than the lattice spacing should agree with CFT results. 
In this section, we will compute the entanglement entropy of $\ell_0$ contiguous sites in the MERA and exploit known CFT results to obtain constraints on the properties of the MERA.
In particular, we will find an inequality relating the MERA rescaling factor $k$ and bond dimension $\chi$ to the CFT central charge $c$.
This constraint is interesting in its own right, but it will prove critical in the next section when we begin to compute bulk properties.

\subsection{MERA and CFT Entanglement Entropy}

For a $(1+1)$-dimensional CFT in a pure state, the von Neumann entropy of a finite interval $B$, which is typically referred to as the entanglement entropy, is known to be \cite{HLWentropy,CardyCalabrese}
\be
S(B) = \frac{c}{3} \ln \ell_0\,,
\label{eq:CCent}
\ee
where the length of the interval is much smaller than the system size.
Here, $\ell_0$ is the length of the interval in units of the UV cutoff.
In the notation of the last section, we have $\ell_0=x_0/a$.
In the special case that the CFT is dual to AdS in $2+1$ dimensions, the central charge is set by the Brown--Henneaux formula \cite{Brown-Henneaux},
\be
c=\frac{3L}{2G}.
\label{brown-henneaux}
\ee
Also note that the length of the geodesic that connects the two ends of $B$ (the curve $\gamma_2$ in \Fig{fig:AdSlines}) is given in \Eq{eq:AdSlengths} by $\left|\gamma_2\right|=2L\ln\ell_0$. 
The Brown--Henneaux relation allows us to reproduce the Ryu--Takayanagi formula \cite{RyuTakayanagi,RyuTakayanagi2} from the entanglement entropy,
\begin{equation}
S(B) = \frac{\mathrm{area}(\tilde B)}{4G} \, ,
\end{equation}
where $\tilde B=\gamma_2$ is the extremal bulk surface with the same boundary as $B$. 
For a boundary with one spatial dimension and a bulk with two spatial dimensions, any simply-connected region $B$ is an interval, the extremal bulk surface is a geodesic, $\mathrm{area}(\tilde B)$ is a length, and $G$ has mass dimension $-1$. 

The MERA calculation of the entanglement entropy of $\ell_0$ sites in the CFT has an analogous geometric interpretation. Suppose one is given the MERA representation of a lattice CFT ground state, \emph{i.e.}, one uses a MERA to generate the CFT state. Denote by $S_\mathrm{MERA}(\ell_0)$ the entanglement entropy of the resulting state restricted to $\ell_0$ sites. In \Ref{Evenbly2014}, it is shown that for a specific, optimal choice of $\ell_0$ sites, for $\ell_0$ parametrically large, the following bound is placed on $S_\mathrm{MERA}(\ell_0)$ for a MERA with $k=2$:
\begin{equation} \label{eq:EVbound}
S_\mathrm{MERA}(\ell_0) \leq 2 \log_2 \ell_0 \ln \chi.
\end{equation}
Parsing the equation above, this bound essentially counts the number of bonds that the \emph{causal cone} of the $\ell_0$ sites in question crosses ($\sim 2 \log_2 \ell_0$) and $\ln \chi$ is the maximum entanglement entropy that a single bond can possess when the rest of the MERA is traced out.

\begin{figure}[ht]
\begin{center}
\includegraphics[scale=0.4]{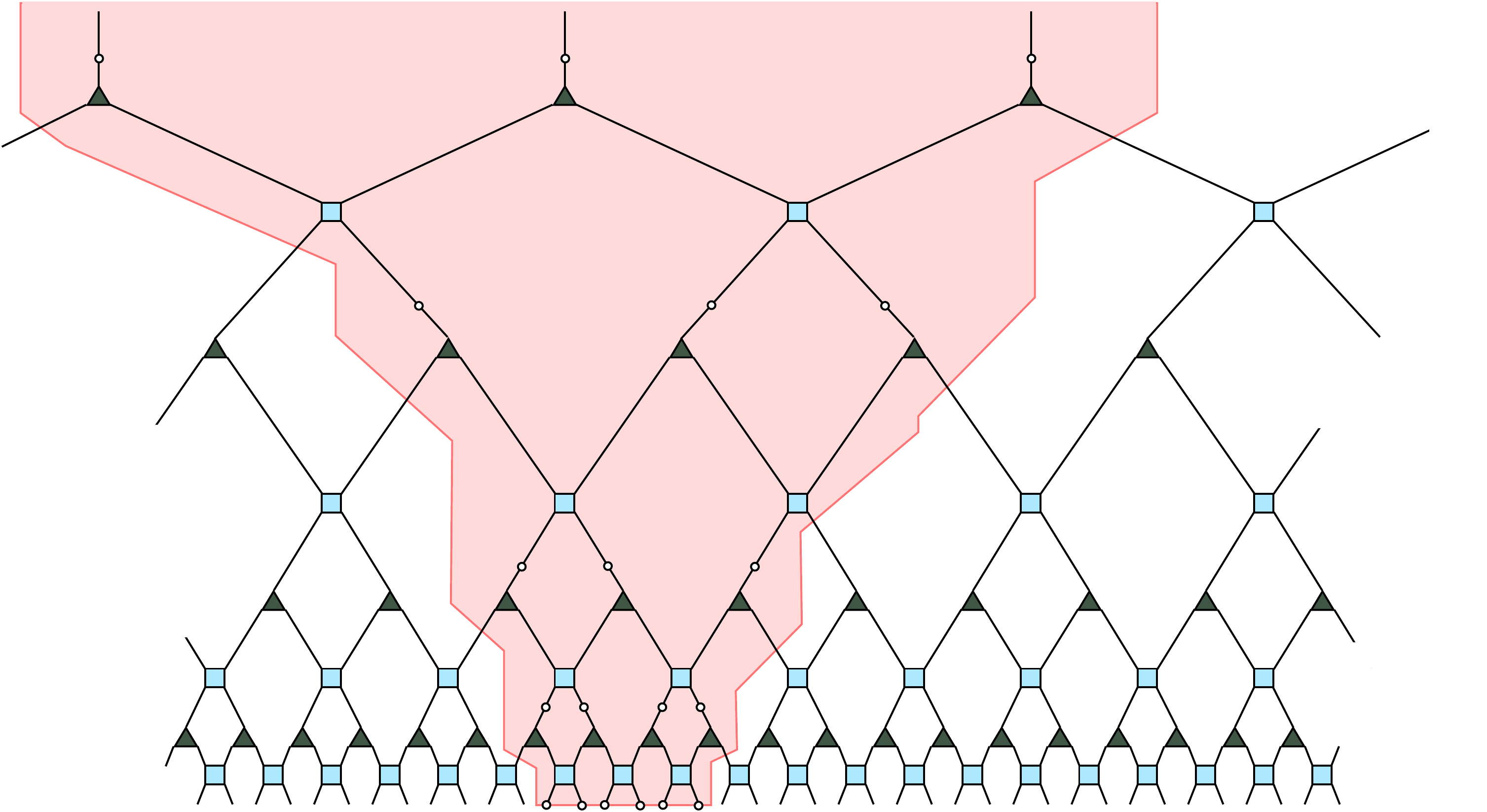}
\caption{Causal cone (shaded) for a set of $\ell_0 = 6$ sites in a MERA with $k=2$. The width $\ell_m$ of the causal cone at depth $m$ is $\ell_1 = 4$, $\ell_2 = 3$, $\ell_3 = 3$, $\ell_4 = 3$, etc. The crossover scale for this causal cone occurs at $\bar m = 2$. Between the zeroth and first layer, $n_1^\mathrm{tr} = 2$ bonds are cut by the causal cone. Similarly, $n_2^\mathrm{tr} = 2$, $n_3^\mathrm{tr} = 3$, etc. }
\label{fig:causalCone}
\end{center}
\end{figure}

The causal cone of a region $B$ consisting of $\ell_0$ contiguous UV sites in a MERA resembles a bulk extremal surface for the boundary region $B$. Given $\ell_0$ sites in the UV, their causal cone is defined as the part of the MERA on which the reduced density matrix (or in other words, the state) of $B$ depends. An example of a causal cone is illustrated in \Fig{fig:causalCone}.

In particular, note that the number of bonds that a causal cone crosses up to any fixed layer scales like the length of the boundary of the causal cone up to that layer. It is in this sense that \Eq{eq:EVbound} is a MERA version of Ryu--Takayanagi. Also note that the width of the causal cone shrinks by a factor of $\sim 1/k$ after every renormalization step until its width is comparable to $k$. As such, if one denotes the width of the causal cone at a layer $m$ by $\ell_m$, then $\ell_m$ is roughly constant for all $m$ greater than some $\bar m$ (see \Fig{fig:causalCone}). The scale $\bar m$ is called the \emph{crossover scale}.

For general $k$, it is also possible to formulate a bound similar to \Eq{eq:EVbound} for the entanglement entropy of $\ell_0$ sites. For parametrically large $\ell_0$, we find that
\begin{equation}  \label{eq:MERAbound}
S_\mathrm{MERA}(\ell_0;B) \leq 4(k-1)\log_k\ell_0 \ln\chi \, . 
\end{equation}
We demonstrate this bound in \App{app:boundProof} using techniques that are similar to those developed in \Ref{Evenbly2014}. In particular, note that we do not allow ourselves to choose the location of the $\ell_0$ sites in question. As such, we remind ourselves that $S_\mathrm{MERA}$ can depend on the location of the region $B$ (and not only its size) by including it in the argument of $S_\mathrm{MERA}$. This is also the reason why our \Eq{eq:MERAbound} is more conservative than the optimal bound given in \Eq{eq:EVbound}.

\subsection{Constraining $S_\mathrm{MERA}$}

Let us examine \Eq{eq:MERAbound} a bit more closely. As discussed in \App{app:boundProof}, $4(k-1)$ is an upper bound on the number of bonds that the causal cone could cut at any given depth $m$ below the crossover scale $\bar m$. (The crossover scale $\bar m$ is attained after roughly $\log_k \ell_0$ renormalization steps.) For a given causal cone, \emph{i.e.}, for  $\ell_0$ sites at a given location with respect to the MERA, let us parametrize our ignorance by writing
\be  \label{eq:MERAbound2}
S_\mathrm{MERA}(\ell_0 ; B) \leq 4 f_B(k) \log_k\ell_0 \ln\chi \, ,
\ee 
where $f_B(k)$ grows no faster than $(k-1)$ and counts the (average) number of bonds cut by the causal cone at any depth up to the crossover scale. Explicitly,
\begin{equation}
f_B(k) \equiv \frac{1}{4 \bar m} \sum_{m=0}^{\bar m - 1} n_m^{\mathrm{tr}} \, ,
\end{equation}
where $n_m^{\mathrm{tr}}$ denotes the number of bonds that the causal cone cuts at the $m^\mathrm{th}$ level.

Each cut bond contributes at most $\ln \chi$ to the entropy (the case of maximal entanglement).  As such, it is instructive to introduce a parameter $\eta_B \in [0,1]$ that describes the degree of entanglement of the sites in the causal cone. In doing so we may rewrite the inequality \eqref{eq:MERAbound2} as an equality:
\be \label{eq:MERAentropy}
S_\mathrm{MERA}(\ell_0;B) = 4 f_B(k)\log_k\ell_0 \cdot \eta_B \ln\chi.
\ee 
The quantity $\eta_B \ln\chi$ is the average entanglement entropy per cut bond in the causal cone of $B$. Equivalently, \Eq{eq:MERAentropy} may be taken as the definition of $\eta_B$.

This definition of $\eta_B$ of course depends on the location of $B$ and only applies to bonds that are cut by the causal cone of $B$. In what follows, it will be advantageous to have a notion of average entanglement entropy per bond that applies to \emph{all} bonds in the MERA. To this end, start with a lattice consisting of $\ell_\mathrm{tot}$ sites in total and consider the limit in which the size of a region $B$ is unbounded but where the ratio $\ell_0 / \ell_\mathrm{tot}$ is held constant (so that $B$ does not grow to encompass the whole domain of the CFT). In this limit, $S_\mathrm{MERA}(\ell_0;B) \rightarrow S_\mathrm{MERA}(\ell_0)$ and $f_B(k) \rightarrow f(k)$ should be independent of the exact location of $B$, \emph{i.e.}, $S_\mathrm{MERA}$ should exactly agree with \Eq{eq:CCent}. Let us consequently define the average entanglement entropy per bond in the MERA:
\be  \label{eq:etadef}
\eta \ln\chi = \lim_{\ell_0 \rightarrow \infty} \, \frac{S_\mathrm{MERA}(\ell_0)}{4f(k)\log_k(\ell_0)} \, ,
\ee 
The quantity $\eta$ is then a property of the MERA itself.

\begin{figure}[ht]
\centering
\subfloat[]{
\includegraphics[scale=0.83,width=0.37\textwidth]{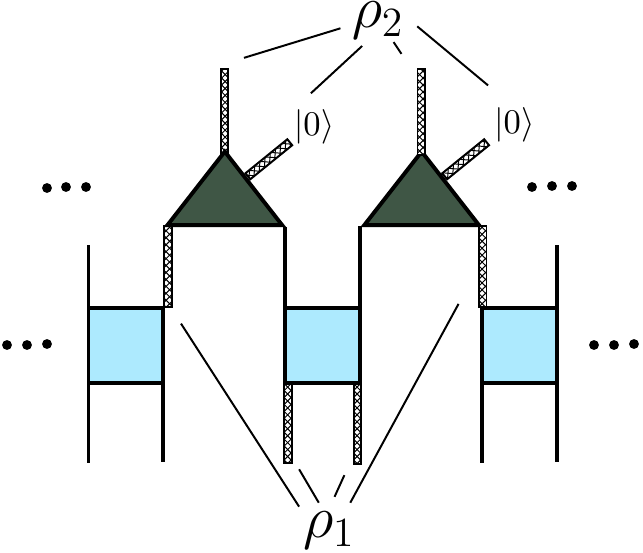}
\label{fig:etabound_2}
}
~~
\subfloat[]{
\includegraphics[scale=0.85,width=0.5\textwidth]{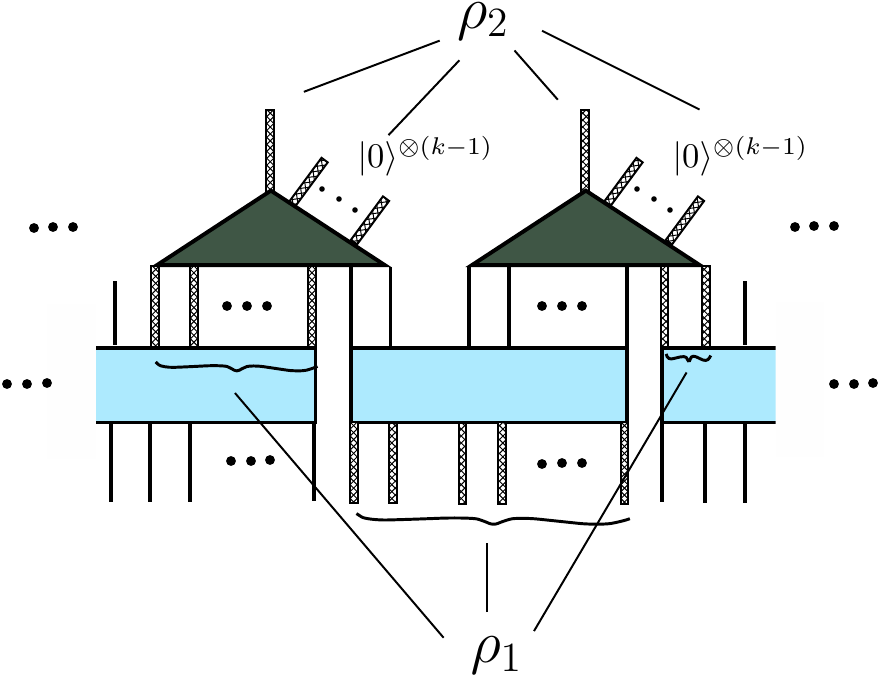}
\label{fig:etabound_k}
}
\caption{A pair of isometries with their ancillae explicitly indicated for a MERA with (a) $k=2$ and (b) general $k$. The thick bonds below the isometries, the state of which is denoted by $\rho_1$, are unitarily related to the bonds that exit the isometries and the ancillae, the state of which is denoted by $\rho_2$. }
\end{figure}

Intuitively, one would not expect each individual bond in the MERA to be maximally entangled and so it should be possible to constrain $\eta$ more tightly than $\eta \leq 1$. This expectation is made more precise via the following considerations. To begin, consider a MERA with $k=2$ and examine a pair of isometries at a fixed depth $m$. As indicated in \Fig{fig:etabound_2}, let $\rho_2$ denote the density matrix of the bonds and ancillae emanating from the two isometries and let $\rho_1$ denote the density matrix of the four highlighted bonds below the isometries. We again assume that the ancillae are initialized to the pure product state composed of factors of $\ket{0}$. Taking into account the ancillae, or in other words promoting the isometries to unitaries, we see that $\rho_1$ and $\rho_2$ are related by a unitary transformation, so $S(\rho_1) = S(\rho_2)$. By assumption, the state of each ancilla is $\ket{0}$, so $\rho_2 = \tilde \rho_2 \otimes \ket{0}\bra{0} \otimes \ket{0}\bra{0}$ for some density matrix $\tilde \rho_2$. This in turn implies that $S(\rho_2) = S(\tilde \rho_2) \leq 2\ln \chi$. From the definition of $\eta$ above, the entanglement entropy of a single bond is asymptotically given by $\eta \ln \chi$, so $S(\rho_1) \simeq 4\eta\ln\chi$. It therefore follows that $\eta \leq 1/2$.

For general $k$, the argument is nearly identical. We again begin by considering a pair of isometries at a given level $m$ (see \Fig{fig:etabound_k}). Analogously with the $k=2$ case, let $\rho_2$ denote the density matrix of the two bonds and $2k-2$ ancillae emanating from the two isometries and let $\rho_1$ denote the density matrix of the $2k$ highlighted bonds below the isometries. There is only one disentangler that straddles both of the isometries in question for any layout of the MERA. As such, at most $k$ of the lower bonds enter a disentangler from below and the rest directly enter the isometries. Here as well $\rho_1$ and $\rho_2$ are related by a unitary transformation so that $S(\rho_1) = S(\rho_2)$. Similarly, $\rho_2 = \tilde \rho_2 \otimes (\ket{0}\bra{0})^{\otimes 2k-2}$ for some density matrix $\tilde \rho_2$, so $S(\rho_2) = S(\tilde \rho_2) \leq 2\ln \chi$. The region described by $\rho_1$ always consists of $2k$ bonds,  so we may again asymptotically write $S(\rho_1) \simeq 2k\eta \ln \chi$. It therefore follows that $k\eta \leq 1$, and since $f(k) \leq (k-1)$, we may write
\begin{equation} \label{eq:eta-k}
\eta f(k) \leq \frac{k-1}{k} \, .
\end{equation}
We note that, in computational practice, one typically does not use the ``worst-case scenario'' construction explored in \App{app:boundProof}; a more conventional construction would result in a tighter bound on $f(k)$ and hence a stricter inequality than \Eq{eq:eta-k}. For our purposes, however, we will remain as conservative as possible and therefore use the inequality \eqref{eq:eta-k} in our subsequent bounds.

\subsection{Matching to the CFT \label{sec:RTmatching}}

Finally, we obtain a constraint on $k$, $\chi$, and $\eta$ in terms of the central charge $c$ by collecting the results of this section. Let us work in the limit where the interval is much larger than the lattice spacing, $\log_k \ell_0 \gg 1$.
We have seen that this is precisely the regime in which $\eta$ and $f(k)$ are well-defined quantities independent of the choice of $B$.
It is also the regime in which we can equate the CFT entropy $S(\ell_0) = (c/3) \ln \ell_0$ with the MERA entropy \eqref{eq:MERAentropy}. Doing so, the central charge is given by
\be 
c=\frac{3L}{2G} = 12\eta \, f(k) \frac{\ln \chi}{\ln k} \, . \label{eq:charge}
\ee
Then in light of \Eq{eq:eta-k}, we find that
\begin{equation}
c \leq 12 \left(\frac{k-1}{k \ln k}\right)\ln \chi \, .\label{eq:chargeconstraint}
\end{equation}

To recapitulate, given a CFT with central charge $c$ and a MERA representation of its ground state, a necessary condition for a consistent AdS/MERA correspondence is that the MERA parameters $k$ and $\chi$ satisfy the constraint \eqref{eq:chargeconstraint}. Importantly, this implies that, for a well-defined semiclassical spacetime (for which $c\gg 1$), the bond dimension $\chi$ must be exponentially large in the size of the AdS scale compared to the Planck scale.

Let us also note that we can still obtain a bound from \Eq{eq:charge}, albeit a weaker one, without using the result of \Eq{eq:eta-k}. Recall that this latter result relies on having unentangled ancillae in the MERA. This is not necessarily the case for other tensor network bulk constructions, as we will subsequently discuss. As such, if we disregard the result of \Eq{eq:eta-k}, we still have by virtue of their definitions that $f(k) \leq k-1$ and $\eta \leq 1$. The following weaker but more general bound on the central charge therefore follows from \Eq{eq:charge} for such generalized tensor networks:
\begin{equation} \label{eq:chargeconstraintweaker}
c \leq 12 \left(\frac{k-1}{\ln k}\right) \ln \chi.
\end{equation}

\section{\label{sec:BH}Constraints from Bulk Entanglement Entropy}

In addition to the compatibility conditions from geodesic matching and boundary entanglement entropy, it is well-motivated to seek out any other possible quantities that can be computed in both the MERA and AdS/CFT frameworks, so as to place further constraints on any AdS/MERA correspondence. One important example of such a quantity is the entropy associated with regions in the bulk, as opposed to on the boundary. 

\subsection{The Bousso Bound}
The notion of placing bounds on the entropy of regions of spacetime in a quantum gravity theory has been explored for many years, first in the context of black hole thermodynamics \cite{Jacobson} and the Bekenstein bound \cite{BekBound} and later in more general holographic contexts, culminating in the covariant entropy bound, \emph{i.e.}, the Bousso bound \cite{BoussoBound1,BoussoBound2}.

The statement of the Bousso bound is the following: given a spacelike surface $\mathcal{B}$ of area $A$, draw the orthogonal null congruence on the surface and choose a direction in which the null generators have non-positive expansion. Let the null geodesics terminate at caustics, singularities, or whenever the expansion becomes positive. The null hypersurface swept out by these null geodesics is called the \emph{lightsheet}. Then the entropy $S$ going through the lightsheet is less than $A/4G$.

Let our spacelike surface $\mathcal{B}$ be a 2-ball of area $A$ on a spacelike slice of AdS and choose as the lightsheet the ingoing future-directed null congruence. This lightsheet will sweep out the entire interior of $\mathcal{B}$ and will terminate at a caustic at the center of $\mathcal{B}$. Since the system is static, the entropy $S$ passing through this lightsheet is the entropy of the system on $\mathcal{B}$, which by the Bousso bound satisfies
\be \label{eq:Bousso}
S(\mathcal{B}) \leq \frac{A}{4G} \, .
\ee 

It is natural to cast the Bousso bound as a constraint on the dimension of the bulk Hilbert space. As argued in \Ref{HolographicPrinciple}, the thermodynamic entropy of a system about which we only know the boundary area $A$ is just the logarithm of the dimension of the true Hilbert space of the bulk region in question (as opposed to the na\"ive Hilbert space in quantum field theory), which the Bousso bound implies is less than $A/4G$.\footnote{\singlespacing\vspace{-10mm}\noindent Moreover, it is known that there exists an asymptotically-AdS bulk configuration that saturates the Bousso bound, namely, the BTZ black hole \cite{BTZ,BTZcorrected}, which further implies that $\ln \dim \mathcal{H}_\mathcal{B}$ in fact equals $A/4G$. However, we will not need this stronger assertion in what follows. A similar but unrelated result equating the area of a region with its entanglement entropy in vacuum was obtained in \Ref{Srednicki}.} As such, if we denote the Hilbert space of $\mathcal{B}$ by $\mathcal{H}_\mathcal{B}$, let us replace \Eq{eq:Bousso} with the slightly more concrete statement
\be \label{eq:Bousso_dim}
\ln \dim \mathcal{H}_\mathcal{B} \leq \frac{A}{4G}.
\ee

\subsection{A MERA version of the Bousso Bound}

Our aim is to compute both sides of the inequality \eqref{eq:Bousso_dim} using the MERA. For this calculation, it is instructive to change our parametrization of the hyperbolic plane from coordinates $(x,z)$, which take values in the half-plane $z > 0$, to coordinates $(\rho,\theta)$, which take values in a disk $0 \leq \rho < 1$, $0 \leq \theta < 2\pi$. Embeddings of the MERA in a disk are often depicted in the literature, \emph{e.g.}, \cite{TNSGeometry}; here we make this coordinate transformation explicit, since we wish to carefully study the geometric properties of the MERA.

\begin{figure}[ht]
\subfloat[]{
\includegraphics[scale=0.5,width=0.38\textwidth]{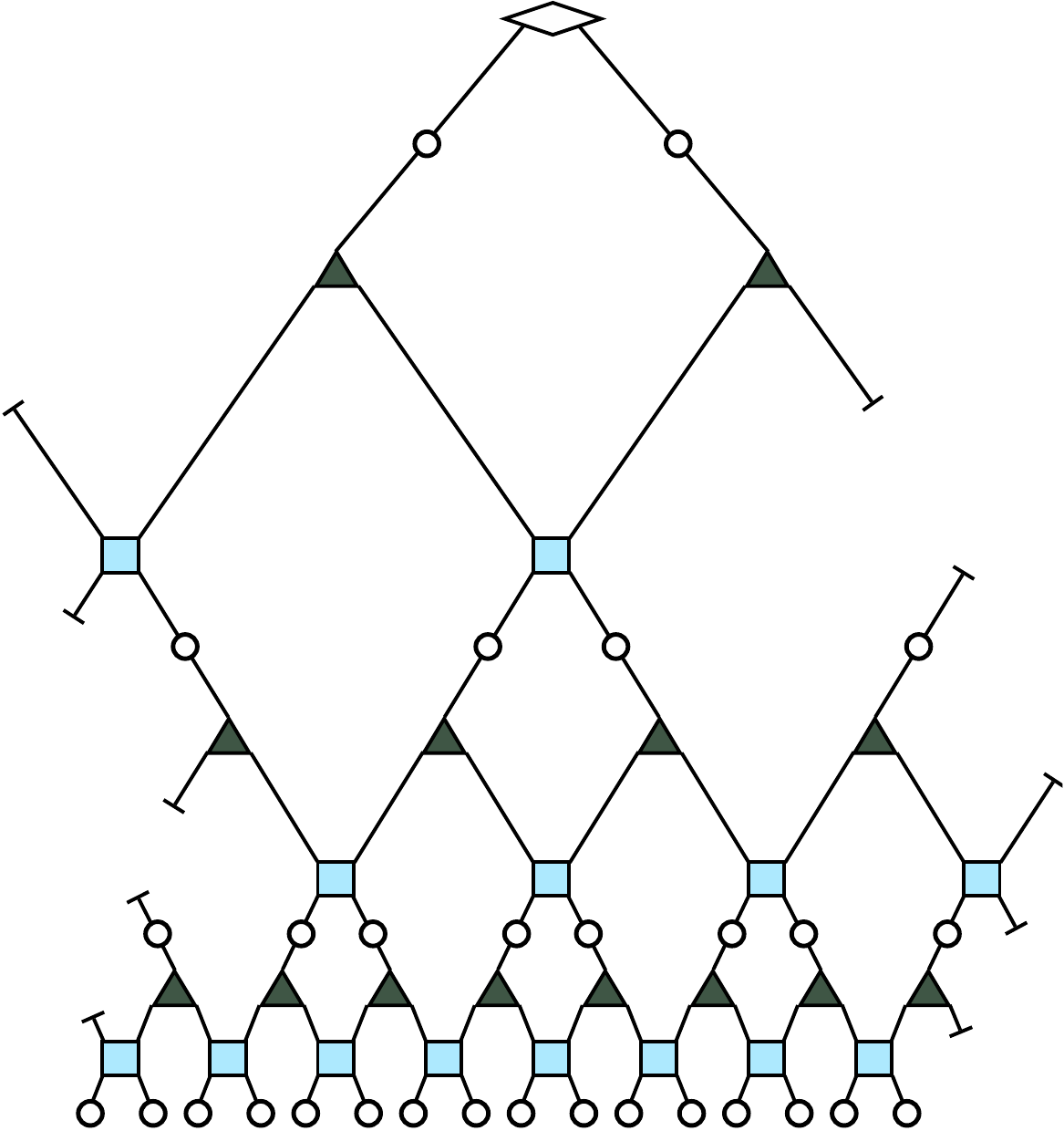}
\label{fig:finiteMERAa}
}
~
\subfloat[]{
\includegraphics[scale=0.74,width=0.42\textwidth]{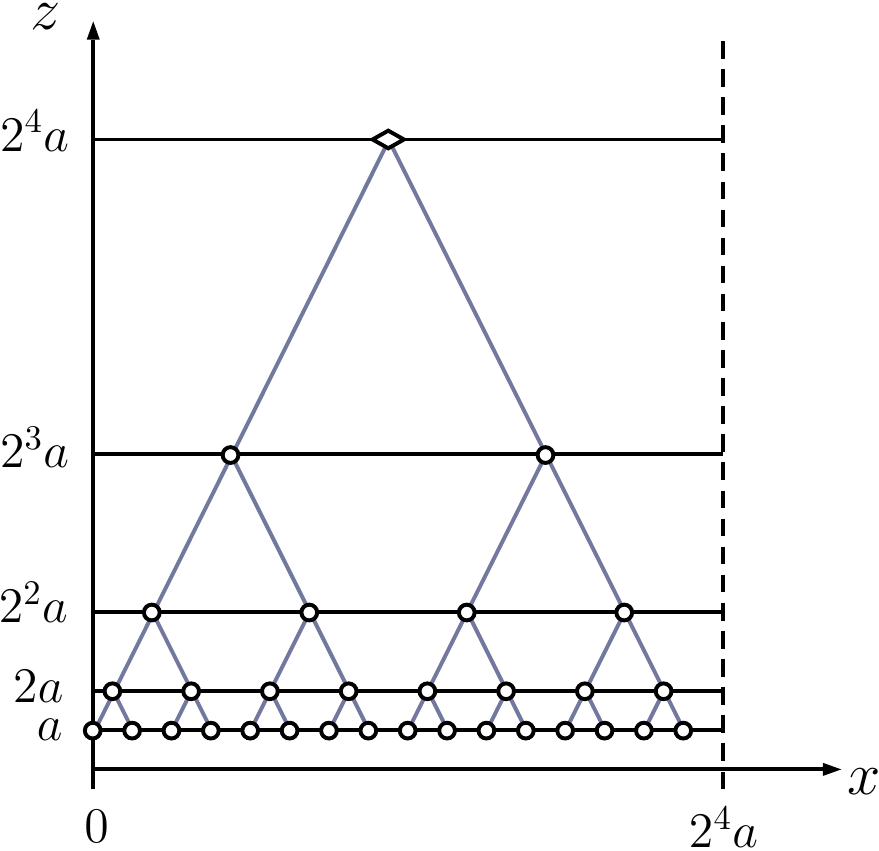}
\label{fig:finiteMERAb}
}
\\
\subfloat[]{
\includegraphics[scale=0.5]{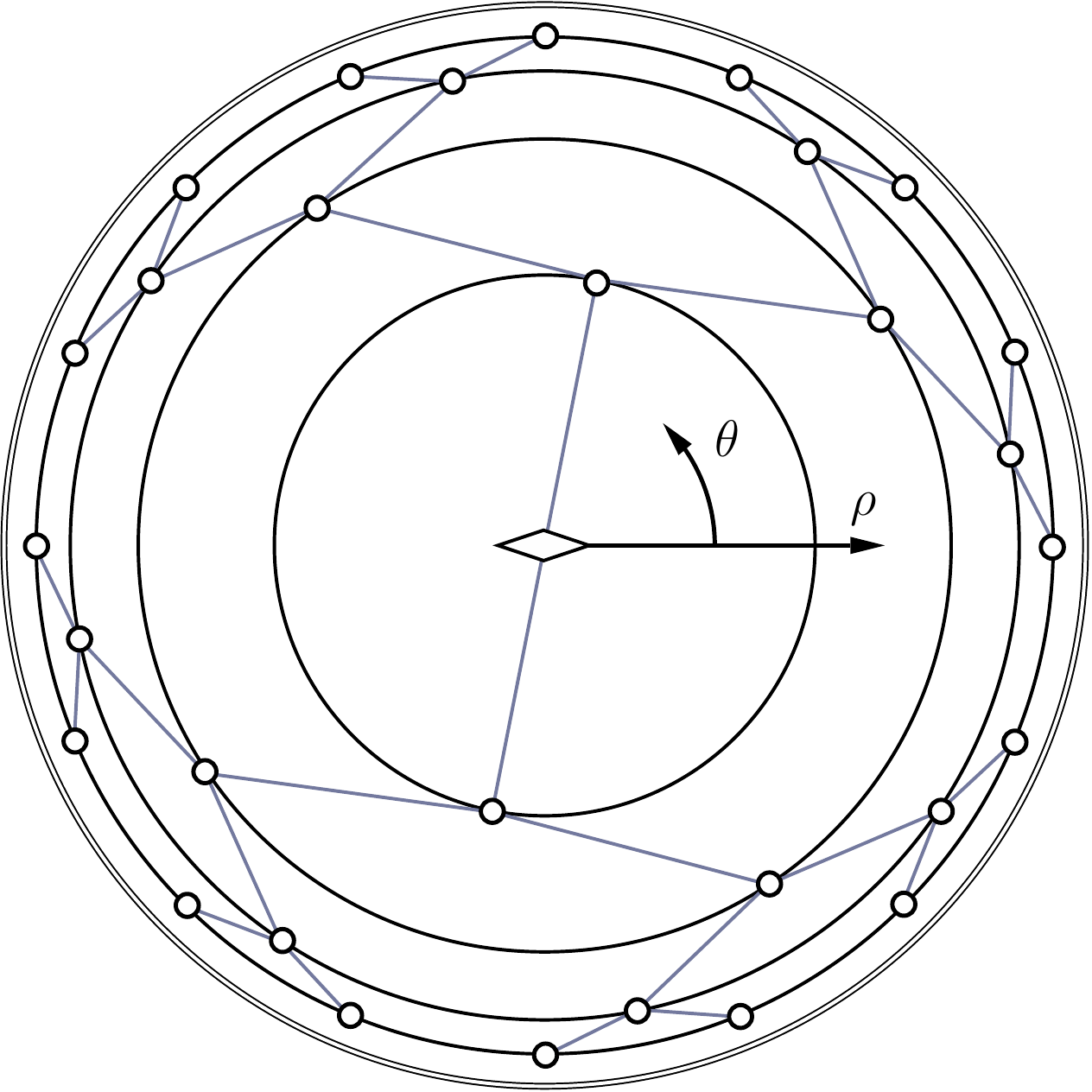}
\label{fig:finiteMERAc}
}
\caption{(a) A $k=2$ MERA consisting of $m=4$ layers and with periodic boundary conditions, (b) the corresponding embedding in $(x,z)$ coordinates, and (c) the embedding  in $(\rho,\theta)$ coordinates. }
\label{fig:finiteMERA}
\end{figure}

To begin, consider a MERA consisting of a single tree that contains a finite number of layers $m$. This situation is illustrated in \Fig{fig:finiteMERAa} for $k=2$ and $m = 4$. Note that such a MERA begins with a top-level tensor at the $m^\mathrm{th}$ level that seeds the rest of the MERA in the IR. 

The base of the MERA is made up of $k^m$ sites. Without loss of generality, let us locate the leftmost site of the base of the MERA at $x = 0$, so that the UV-most sites sit at coordinates $(x,z) = (na,a)$, where $n = 0,1,2, \ldots, (k^m-1)$ as shown in \Fig{fig:finiteMERAb}. Let us also assume periodic boundary conditions for this MERA and hence identify $x=0$ and $x=k^m a$.

Next, define the coordinates $(\rho,\theta)$ as follows:
\begin{equation}
\begin{aligned}
\rho &= \frac{k^m a - z}{k^m a},\\
\theta &= 2 \pi \frac{x}{k^m a}.
\end{aligned}
\end{equation}
In these coordinates, the metric reads
\begin{equation} \label{eq:circularMetric}
\mathrm{d}s^2 = \frac{L^2}{(1-\rho)^2}\left[\mathrm{d}\rho^2 + \left(\frac{\mathrm{d}\theta}{2\pi}\right)^2\right] \, ,
\end{equation}
\emph{cf}. \Eq{eq:AdSmetric}. This embedding of the MERA is shown in \Fig{fig:finiteMERAc}; the top-level tensor always sits at $\rho = 0$ and the lower layers of the MERA are equally spaced on circles of radii $1/2, \, 3/4, \, 7/8, \, \ldots$ that are centered at $\rho = 0$.

\begin{figure}[ht]
\centering
\includegraphics[scale=0.5]{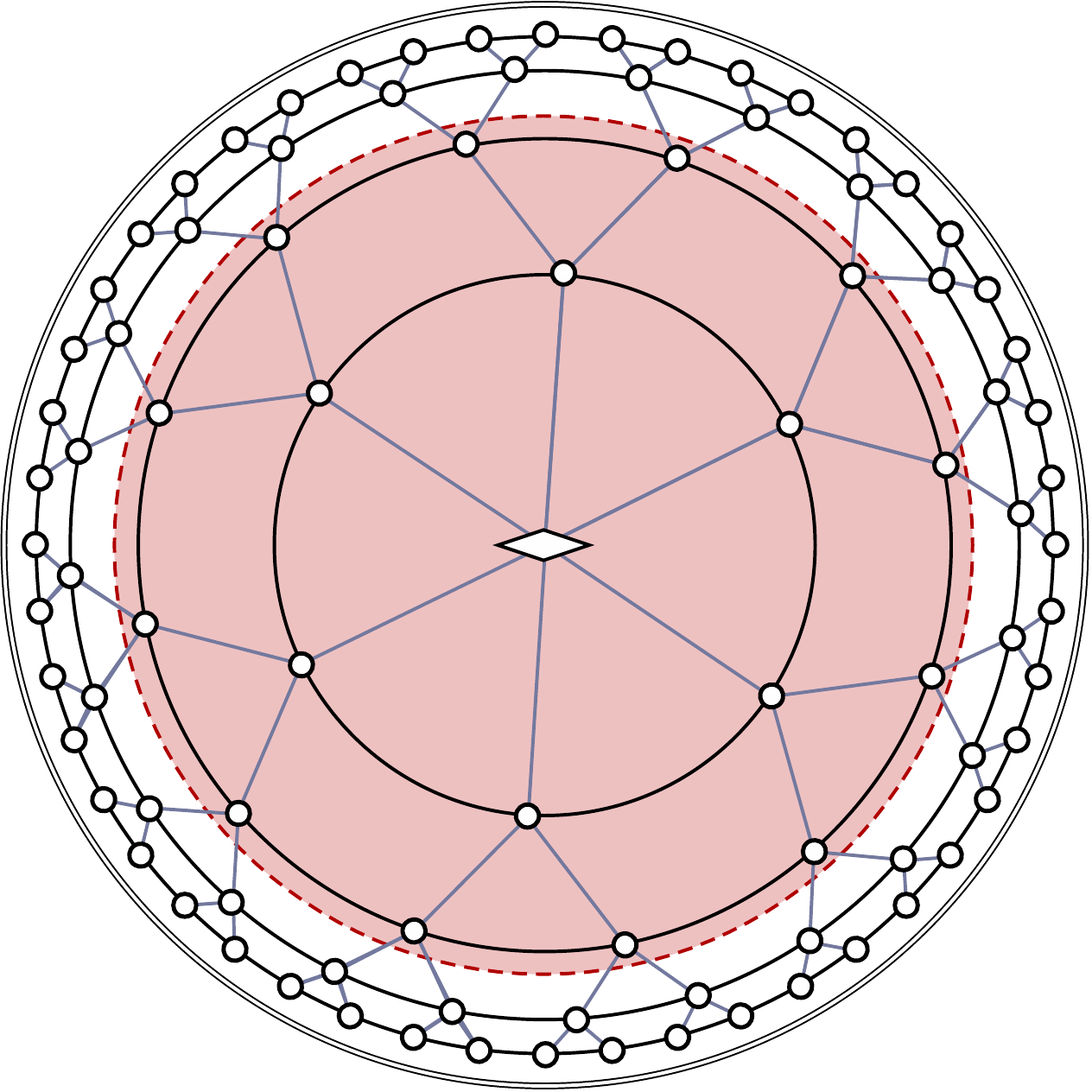}
\caption{Disk parametrization of the Poincar\'e patch of AdS in which a MERA has been embedded. The top tensor of the MERA shown has $T=6$. The shaded region is a ball $\mathcal{B}$, which is this case contains $N_\mathcal{B}=1$ generation. }
\label{fig:topTMERA}
\end{figure}

More generally, one could construct a top-level tensor that has $T$ legs, each of which begets a tree of sites. In this case, $x=0$ and $x=Tk^{m-1}a$ are identified, so one should define the angular variable as $\theta \equiv 2\pi x/(Tk^{m-1}a)$. The metric \eqref{eq:circularMetric} is correspondingly modified and reads
\begin{equation}
\mathrm{d}s^2 = \frac{L^2}{(1-\rho)^2}\left[\mathrm{d}\rho^2 + \frac{T^2}{k^2}\left(\frac{\mathrm{d}\theta}{2\pi}\right)^2\right] \, .
\end{equation} This situation is depicted in \Fig{fig:topTMERA}. (If $T=k$, however, then it is not necessary to introduce any new structure in addition to the disentanglers and isometries that were already discussed, \emph{i.e.}, one may take the top-level tensor to be one of the isometries.)

We may immediately compute the right-hand side of \Eq{eq:Bousso_dim}. Let the ball $\mathcal{B}$ be centered about $\rho = 0$, and suppose $\mathcal{B}$ contains the top-level tensor, the sites at the top tensor's legs, and then the first $N_\mathcal{B}$ generations of the MERA emanating from these sites, as indicated in \Fig{fig:topTMERA}. The boundary of $\mathcal{B}$ is a circle at constant $\rho$, so its circumference according to the MERA is $A = Tk^{N_\mathcal{B}}L$. As such, we may write
\be \label{eq:bulkPerimeter}
\frac{A}{4G} = \frac{Tk^{N_\mathcal{B}}L}{4G} = \frac{Tk^{N_\mathcal{B}} c}{6} \, ,
\ee 
where in the second equality we used the Brown-Henneaux relation, \Eq{brown-henneaux}.

How one evaluates the left-hand side of \Eq{eq:Bousso_dim} using the MERA is not as immediate. Recall that $\mathcal{H}_\mathcal{B}$ is the Hilbert space of \emph{bulk states}. The MERA, however, does not directly prescribe the quantum-gravitational state in the bulk; it is not by itself a bulk-boundary dictionary. As we mentioned in \Sec{sec:AdSMERAcor}, the minimal assumption that one can make is to posit the existence of a bulk Hilbert space factor $V_\mathrm{bulk}$ associated with each MERA site that is not located at the top tensor. To keep the assignment general, we assign a factor $V_\mathrm{T}$ to the top tensor. The dimensionality of each $V_\mathrm{bulk}$ factor should be the same in order to be consistent with the symmetries of the hyperbolic plane. The assumption of a Hilbert space factor at every MERA site is minimal in the sense that it introduces no new structure into the MERA. A true AdS/MERA correspondence should dictate how states in the bulk Hilbert space are related to boundary states. However, for our analysis, it is enough to simply postulate the existence of the bulk Hilbert space factors $V_\mathrm{bulk}$ and $V_\mathrm{T}$, each of which may be thought of as localized to an AdS-scale patch corresponding to the associated MERA site. 

In addition to the site at the top tensor, the number of regular MERA sites that the ball $\mathcal{B}$ contains is given by
\be 
\mathcal{N}_\mathcal{B} = T \sum_{i=0}^{N_\mathcal{B}} k^i = T\left(\frac{k^{N_\mathcal{B}+1}-1}{k-1}\right)  \, \label{eq:scrN}.
\ee
As such, the Hilbert space of bulk states restricted to $\mathcal{B}$ is $\mathcal{H}_\mathcal{B} = (V_\mathrm{bulk})^{\otimes \mathcal{N}_\mathcal{B}}\otimes V_\mathrm{T}$. Next, suppose that $\dim V_\mathrm{bulk} = \tilde \chi$ and that $\dim V_\mathrm{T}=\tilde{\chi}_\mathrm{T}$, where, like $\chi$, $\tilde \chi$ and $\tilde{\chi}_\mathrm{T}$ are some fixed, $N_\mathcal{B}$-independent numbers. Then $\dim \mathcal{H}_\mathcal{B} = \tilde{\chi}_\mathrm{T}(\tilde{\chi}^{\mathcal{N}_\mathcal{B}})$. 
Note that one would expect $\chi$ and $\tilde{\chi}$ to have a very specific relationship in a true bulk/boundary correspondence, the nature of which will be explored later in this section. Combining \Eqs{eq:bulkPerimeter}{eq:scrN}, the dimensionality of $\mathcal{H}_\mathcal{B}$ is upper bounded as follows:
\be 
\ln \dim \mathcal{H}_\mathcal{B} \leq \frac{A}{4G} \quad \implies \quad  T\left(\frac{k^{N_\mathcal{B}+1}-1}{k-1}\right) \ln \tilde \chi + \ln\tilde{\chi}_\mathrm{T}\leq \frac{T k^N_\mathcal{B} c}{6} \, .\label{eq:logdimH}
\ee 
After isolating $c$ in \Eq{eq:logdimH} and using the result of \Eq{eq:charge}, we find that
\be  \label{eq:protoBousso}
c = 12\eta f(k) \frac{\ln \chi}{\ln k} \geq 6 \left(\frac{k^{N_\mathcal{B}+1}-1}{k^{N_\mathcal{B}}(k-1)}\ln \tilde \chi + \frac{1}{Tk^{N_\mathcal{B}}}\ln\tilde{\chi}_\mathrm{T}\right)  \, .
\ee 

Next, let us consider this inequality in the limit of large $N_\mathcal{B}$. A motivation for this limit is the fact that the natural scale of validity of an AdS/MERA correspondence is super-AdS, as was established in \Sec{sec:geodesics}. Moreover, by virtue of its definition, there is always an ambiguity of order the AdS scale in the radius of the ball $\mathcal{B}$. That is, the region in AdS denoted by $\mathcal{B}$ is only well-defined in the MERA if $\mathcal{B}$ is large compared to the AdS scale $L$. Taking the limit of large $N_\mathcal{B}$, \Eq{eq:protoBousso} reduces to 
\be 
\eta f(k) \geq \frac{k \ln k}{2(k-1)} \left( \frac{\ln \tilde \chi}{\ln \chi} \right) \, .\label{eq:protoBoussolargeN}
\ee 
By using the bound on $\eta f(k)$ given by \Eq{eq:eta-k}, we arrive at a constraint on $k$, $\chi$, and $\tilde \chi$:
\be 
\frac{k^2 \ln k}{2(k-1)^2} \left( \frac{\ln \tilde \chi}{\ln \chi} \right) \leq 1. \label{eq:protobound}
\ee 

In principle, the above inequality could be satisfied for any $k$, provided that the dimension $\tilde \chi$ of the factors $V_{\text{bulk}}$ can be arbitrarily chosen with respect to the bond dimension $\chi$. However, the essence of holography, that the bulk and boundary are dual descriptions of the same degrees of freedom and therefore have isomorphic Hilbert spaces \cite{MAGOO}, implies a relation between $\chi$ and $\tilde{\chi}$. Namely, for a MERA with a total of $N$ levels of sites in the bulk strictly between the UV-most level and the top-level tensor, the number of bulk sites $\mathcal{N}_\mathrm{bulk}$ that are not located at the top tensor is given by \Eq{eq:scrN} with $N_\mathcal{B} = N$, and the number of sites in the boundary description is $\mathcal{N}_\mathrm{boundary} \equiv Tk^{N+1}$.
The bulk Hilbert space thus has dimension $\tilde{\chi}^{\mathcal{N}_\mathrm{bulk}}\tilde{\chi}_{\mathrm{T}}$ and the boundary Hilbert space has dimension $\chi^{\mathcal{N}_\mathrm{boundary}}$. Equating\footnote{We recognize that there are other proposals \cite{HQECCs,QEC14} that do not require an exact equivalence between the bulk and boundary Hilbert spaces, but, even in these cases, there is the requirement of an exact equivalence between the logical qubits on the boundary with the Hilbert space of the bulk.} the dimension of the bulk and boundary Hilbert spaces then yields
\be
\frac{\ln \tilde{\chi}}{\ln \chi} = \frac{1}{\mathcal{N}_\mathrm{bulk}}\left(Tk^{N+1}-\frac{\ln\tilde{\chi}_\mathrm{T}}{\ln\chi}\right)\stackrel{N\,\text{large}}{\rightarrow}k-1, \label{eq:chichirel}
\ee
where we took the limit of $N$ large, consistent with \Eq{eq:protoBoussolargeN} and in keeping with the expectation that the UV cutoff be parametrically close to the boundary at $\rho=1$. Putting together \Eqs{eq:protobound}{eq:chichirel}, we obtain a constraint on $k$ alone:
\be \label{eq:BoussoTight}
\frac{k^2 \ln k}{2(k-1)} \leq  1.
\ee 
This constraint cannot be satisfied for any allowed value of the rescaling factor $k$, which must be an integer greater than or equal to 2. We thus learn that a conventional MERA cannot yield a consistent AdS/MERA correspondence. The MERA cannot simultaneously reproduce AdS geodesics, respect the Ryu--Takayanagi relation, and (using the only construction for the bulk Hilbert space available to the MERA by itself) satisfy the Bousso bound. That is, there exists no choice of MERA parameters that can faithfully reproduce geometry, holographic properties, and bulk physics.

If we relax this bound and, instead of \Eq{eq:eta-k}, only observe the weaker, natural bounds $\eta \leq 1$ and $f(k) \leq k-1$ as discussed at the end of \Sec{sec:RTmatching}, the constraint \eqref{eq:BoussoTight} is correspondingly modified:
\begin{equation} \label{eq:BoussoLoose}
\frac{k \ln k}{2(k-1)} \leq 1.
\end{equation}
In contrast to \Eq{eq:BoussoTight}, this latter bound can be satisfied, but only for $k=2$, $3$, or $4$. As such, other AdS/tensor network correspondences, in which the ancillae are perhaps entangled and therefore do not describe a conventional MERA, are not ruled out. Note that we never needed to compute bulk entanglement entropy explicitly --- and therefore did not need to treat separately the possibility of entanglement among ancillae --- because we cast the Bousso bound as a constraint on the size of the bulk Hilbert space itself. The appearance of $\eta$ in \Eq{eq:protoBoussolargeN} corresponds to entanglement in the boundary theory as computed by the tensor network; \Eqs{eq:protoBoussolargeN}{eq:chichirel} still apply.

\section{\label{sec:conc}Conclusion}

The notion of emergence of spacetime based on a correspondence between AdS and a tensor network akin to AdS/CFT is a tantalizing one. A necessary step in such a program is the evaluation and comparison of calculable quantities on both sides of the duality. In this work, we have subjected the proposed AdS/MERA correspondence to such scrutiny. To summarize, let us restate our three main findings:
\begin{enumerate}
\item In matching the discrete graph geometry of the MERA to the continuous geometry of a spatial slice of AdS, we demonstrated that the MERA describes geometry only on scales larger than the AdS radius. Concretely, as shown in \Sec{sec:geodesics}, the proper length assigned to the spacing between adjacent sites in the MERA lattice must be the AdS scale.
\item By requiring that the entropy of a set of boundary sites in the MERA --- whose computation is a discrete realization of the Ryu--Takayanagi formula --- be equal to the CFT ground state entropy of the same boundary region in the thermodynamic limit, we obtained a constraint on the parameters that describe a MERA in terms of the CFT central charge [\Eqs{eq:chargeconstraint}{eq:chargeconstraintweaker}], which implies that the bond dimension $\chi$ must be exponentially large in the ratio of the AdS scale to the Planck scale.
\item In the natural construction of a bulk Hilbert space  ($\mathcal{H}_\text{bulk}$) using the MERA, we used the Bousso bound to constrain the dimension of $\mathcal{H}_\text{bulk}$. When combined with our previous results, we found that any strict AdS/MERA correspondence cannot satisfy the resulting constraint, \Eq{eq:BoussoTight}. Upon relaxing the definition of the MERA or allowing for additional structure, however, we obtained a looser constraint, \Eq{eq:BoussoLoose}, which may not rule out some other AdS/tensor network correspondences.
\end{enumerate}
In particular, more general correspondences between AdS and MERA-like tensor networks, in which we allow the ancillae to be entangled when reproducing the CFT ground state [and for which \Eq{eq:BoussoLoose} applies in place of \Eq{eq:BoussoTight}] are not ruled out by our bounds, provided that the rescaling factor $k=2$, $3$, or $4$. Further, it is interesting to note that our bounds extend to states other than the vacuum that are described by a MERA. One such example, namely, states at finite temperature dual to black holes in AdS, is discussed in \App{appBTZ} below.

While the consistency conditions that we found are specific to the MERA tensor network, many of the ideas and techniques that we used apply equally well to other tensor networks. In the EHM, for instance, the type of bulk Hilbert space dimensionality arguments that we made based on the covariant entropy bound may be directly transferred to the EHM. The same stringent final constraints that we derived do not apply to the EHM, however, since it is unclear to what extent the EHM reproduces the Ryu--Takayanagi formula (which renders the results of \Sec{sec:entanglement} inapplicable). Our bulk Hilbert space arguments similarly apply to the holographic error-correcting code proposal in \Ref{HQECCs}, which furthermore purports to reproduce a version of the Ryu--Takayanagi formula. It is presently unknown, however, whether the boundary state of a holographic code can represent the ground state of a CFT, so an identification of entropies similar to the identification $S_\mathrm{MERA} = S_\mathrm{CFT}$, upon which our boundary entropy constraints so crucially depend, cannot yet be made.

In closing, we have found several consistency conditions that any AdS/MERA correspondence must satisfy. The totality of these constraints rules out the most straightforward construal of an AdS/MERA correspondence. Other interesting holographic correspondences that are described by tensor networks more general than the MERA and that respect all of our bounds may indeed be possible. Our consistency conditions are nice validity checks for these correspondences when applicable and in other cases they may inspire similar consistency conditions. The program of identifying the emergence of spacetime from the building blocks of quantum information is an ambitious one; stringent consistency conditions, such as those presented in this paper, are important for elucidating the subtleties in this quest and in providing guidance along the way.

\begin{center} 
 {\bf\small ACKNOWLEDGMENTS}
 \vspace{-2mm}
 \end{center}
 \noindent  We thank Bartek Czech, Glen Evenbly, Daniel Harlow, Shamit Kachru, Aleksander Kubica, Shaun Maguire, Spiros Michalakis, Don Page, John Preskill, Bogdan Stoica, James Sully, Brian Swingle, and Guifr\'e Vidal for helpful discussions.
This research was supported in part by DOE grant DE-SC0011632 and by the Gordon and Betty Moore Foundation through Grant 776 to the Caltech Moore Center for Theoretical Cosmology and Physics. N.B. is supported by the DuBridge postdoctoral fellowship at the Walter Burke Institute for Theoretical Physics. A.C.-D. and C.C. are supported by the NSERC Postgraduate Scholarship program. G.N.R. is supported by a Hertz Graduate Fellowship and a NSF Graduate Research Fellowship under Grant No.~DGE-1144469.

\appendix

\section{Entropy bound for general MERAs \label{app:boundProof}}

Following the method presented in \Ref{Evenbly2014}, let us compute an upper bound for the entanglement entropy of a region $B$ consisting of $\ell_0$ sites in a MERA with rescaling factor $k$. We will use the notation of \Ref{Evenbly2014} throughout. 

First, recall the result from \Ref{Evenbly2014} that the entanglement entropy of a region consisting of $\ell_0$ sites is bounded by
\be  \label{eq:generalSbound}
S_\mathrm{MERA}(\ell_0 ; B)\leq (\ell_{m'}+N_{m'}^\mathrm{tr})\ln \chi.
\ee 
The quantity $\ell_{m'}$ is the width of the causal cone at depth $m'$ and $N_{m'}^\mathrm{tr} =\sum_{m=0}^{m'-1} n_m^\mathrm{tr}$ is the total number of sites that are traced out along the boundary of the causal cone. In other words, $N_{m'}^\mathrm{tr}$ is the number of bonds that are cut by the causal cone up to a depth $m^\pr$ (\emph{cf}. \Fig{fig:causalCone}). The quantity $\ln \chi$ is the maximum entanglement entropy that each site that is traced out could contribute to $S_\mathrm{MERA}(\ell_0 ; B)$. Note that \Eq{eq:generalSbound} holds for all $m^\pr \geq 0$.

The width of the causal cone for a given $m^\pr$ depends sensitively on the structure of the MERA. In particular, the number of sites that are traced out at each renormalization step depends on the choice of disentanglers, as well as how they are connected to the isometries. For instance, in a MERA with a rescaling factor $k$, any given disentangler could have anywhere from 2 up to $k$ incoming and outgoing legs. (It should be reasonable to require that any disentangler can have no more than $k$ incoming and $k$ outgoing legs so that it straddles no more than two isometries.) It is thus clear that the number of bonds that one cuts when drawing a causal cone, and hence the entanglement entropy of the region subtended by that causal cone, depends on the choice of disentanglers and connectivity.

\begin{figure}[ht]
\begin{center}
\includegraphics[scale=1.25]{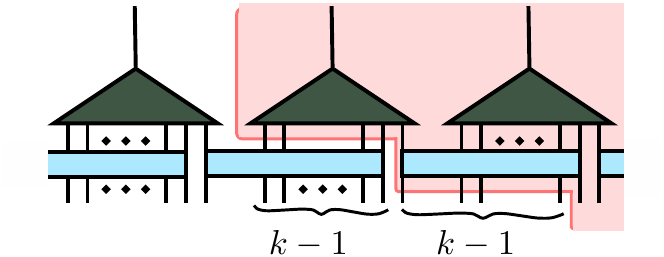}
\caption{Left side of a causal cone that cuts the maximum possible number of bonds over the course of one renormalization step. The rectangles are disentanglers that accept $k$ bonds as input and the triangles are isometries that coarse-grain $k$ bonds into one. The causal cone is the shaded region. If this situation is mirrored on the right side of the causal cone, then $4(k-1)$ bonds are cut in this renormalization step. }
\label{fig:kto1bound}
\end{center}
\end{figure}

Nevertheless, we can compute an upper bound for $S_\mathrm{MERA}(\ell_0 ; B)$ by considering a worst-case scenario for the number of bonds cut by the causal cone. We begin by asking: What is the largest number of bonds that a causal cone could cut in one renormalization step at a depth $m^\pr$? The layout of disentanglers and isometries that produces this situation is shown at one side of a causal cone in \Fig{fig:kto1bound}. If the causal cone at the bottom of the renormalization step incorporates a single bond that goes into a disentangler accepting $k$ bonds, then the causal cone must cut the other $k-1$ bonds entering the disentangler. Then if this disentangler is arranged so that its leftmost outgoing bond is the first bond to enter an isometry from the right, the causal cone must cut the other $k-1$ bonds entering the isometry. If this arrangement is mirrored on the other side of the causal cone, we see that $4(k-1)$ bonds are cut by the causal cone in this renormalization step, \emph{i.e.}, $n_{m'}^\mathrm{tr}=4(k-1)$.

Recall that for any finite $\ell_0$, after a fixed number of renormalization steps, the width of the causal cone remains constant for any further coarse-grainings. The depth at which this occurs is called the crossover scale and is denoted by $\bar m$. Therefore, the causal cone will cut the largest possible number of bonds when the arrangement described above and depicted in \Fig{fig:kto1bound} occurs at every step up until the crossover scale.
Then, by \Eq{eq:generalSbound}, the entropy bound is given by 
\be
S_\mathrm{MERA}(\ell_0 ; B) \leq (\ell_{\bar{m}}+4(k-1)\bar{m})\ln\chi,
\label{eq:Sbound2}
\ee
where $\ell_{\bar{m}}$ is the width of the causal cone at the crossover scale.

For any given causal cone in a MERA with scale factor $k\geq2$, the maximum number of additional sites the causal cone can pick up at some level $m'$ is $4(k-1)$. Therefore, for a causal cone that contains $\ell_{m'}$ sites at depth $m'$, the number of sites in the causal cone after one renormalization step $\ell_{m'+1}\leq \ceil*{(\ell_{m'}+4(k-1))/k}\leq \ell_{m'}/k+5$. Applying the relation recursively, we find that the number of sites $\ell_{m'}$ at any layer $m^\prime < \bar{m}$ is bounded, 
\be 
\ell_{m^\prime} ~ \leq ~ \frac{\ell_0}{k^{m^\prime}}+5\sum_{m=1}^{m^\prime} \frac{1}{k^m} ~ \leq ~ \frac{\ell_0}{k^{m^\prime}}+5 \, .
\ee 
Setting $m'=\bar{m}$, it trivially follows that the crossover scale obeys $\bar{m}\leq\log_k \ell_0$. 
Furthermore, we notice that this is the scale at which the entanglement entropy is minimized if we trace over the remaining sites. In other words, the  number of bonds cut by going deeper into the renormalization direction is no less than the bonds cut horizontally, so $4(k-1)\geq \ell_{\bar{m}}$ \footnote{\singlespacing\vspace{-10mm}\noindent Alternatively, we can see this from a heuristic argument by noting that the crossover scale is the scale at which the causal cone has a constant width for further coarse-grainings, \emph{i.e.}, $(\ell_{\bar{m}}+4(k-1))/k\approx\ell_{\bar{m}}$. Therefore, $\ell_{\bar{m}}\lesssim 4\leq 4(k-1)$.}. Applying the bounds for $\bar{m}$ and $\ell_{\bar{m}}$ on \Eq{eq:Sbound2}, we arrive at an upper bound on $S_\mathrm{MERA}(\ell_0 ; B)$  for a $k$-to-one MERA,
\be 
S_\mathrm{MERA}(\ell_0 ; B) \leq 4(k-1)(1+\log_k \ell_0)\ln\chi.
\ee
When $\ell_0$ is parametrically large, we neglect the $\mathcal{O}(1)$ contribution to the bound on $S_\mathrm{MERA}(\ell_0 ; B)$, which yields \Eq{eq:MERAbound}.

\section{BTZ Black Holes and Thermal States in AdS/MERA}
\label{appBTZ}
Thus far, we have found constraints on the structure of a MERA that can describe CFT states dual to the AdS$_3$ vacuum. One might ask whether these results extend to other constructions that exist in three-dimensional gravity. Although pure gravity in AdS$_3$ has no local or propagating degrees of freedom, there exist interesting non-perturbative objects, namely, BTZ black holes \cite{BTZ}. In this appendix, we extend our constraints on boundary entanglement entropy to these objects.

The non-rotating, uncharged BTZ black hole solution is given in Schwarzschild coordinates by
\be 
{\rm d}s^2 = - \frac{(r^2 - r^2_+)}{L^2} {\rm d}t^2 + \frac{L^2}{(r^2 - r^2_+)} {\rm d}r^2 + r^2 {\rm d}\phi^2\,,
\label{eq:BTZmetric}
\ee 
with a horizon at $r=r_+$.
Noting that Euclidean time is compactified by identifying $\tau \sim \tau +2\pi L^2/r_+$, the horizon temperature of the  black hole is given by $T=r_+/2\pi L^2$. Additionally, the Bekenstein--Hawking entropy of the black hole is
\be 
S_{\rm BH} = \frac{{\rm Area}}{4G} = \frac{\pi r_+}{2G}\,. \label{eq:BTZentropy}
\ee 

\begin{figure}[ht]
\begin{center}
\includegraphics[scale=0.5]{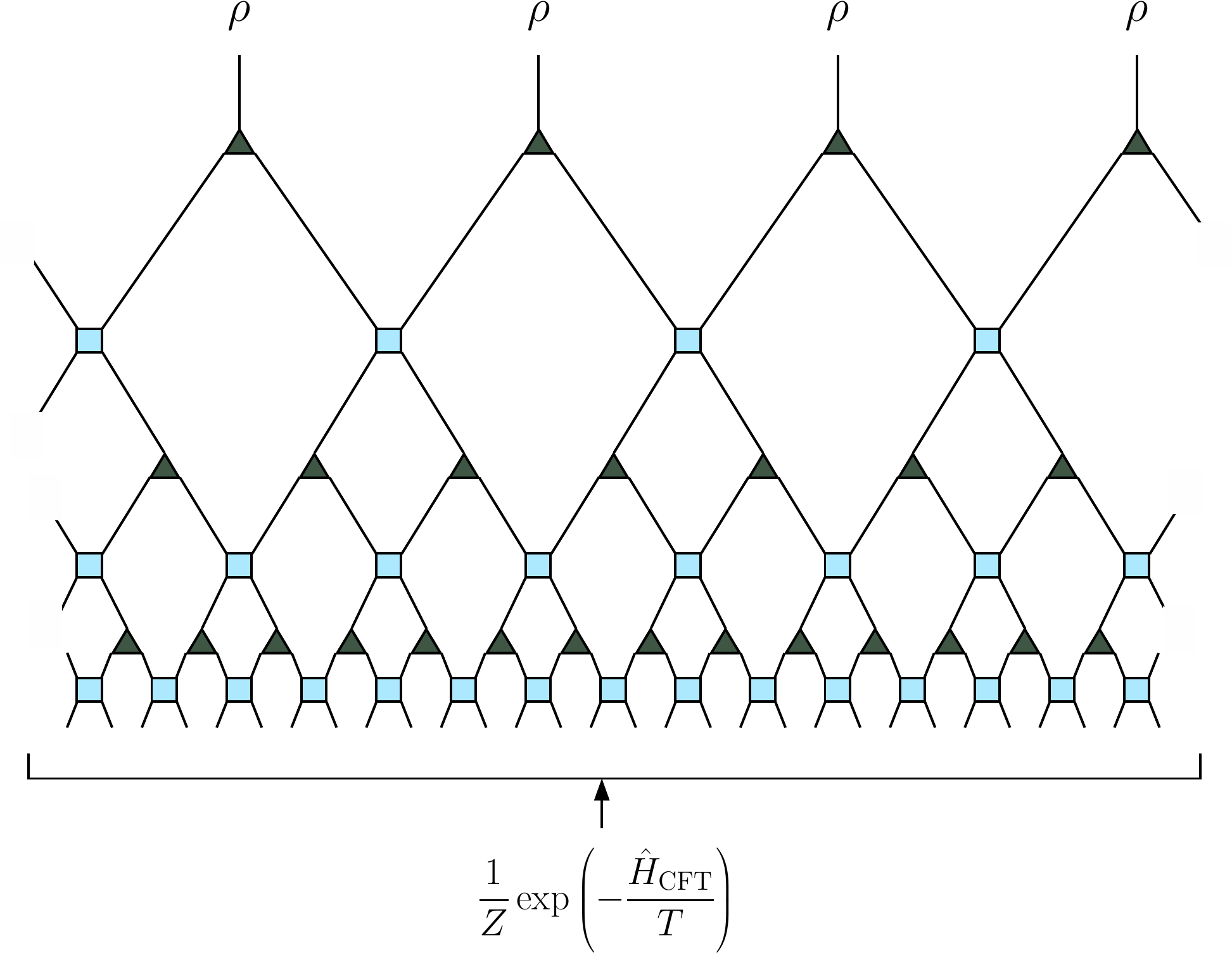}
\caption{The MERA, when applied to a thermal CFT state $Z^{-1} \exp(-{\hat H}_\mathrm{CFT}/ T)$, where $Z = \mathrm{tr}(\exp(-{\hat H}_\mathrm{CFT}/T))$, truncates after a finite number of layers. The boundary state at the top of the truncated MERA effectively factorizes into a product of maximally mixed states $\rho = I/\chi$.}
\label{fig:thermalMERA}
\end{center}
\end{figure}

Let us now consider applying a MERA with rescaling factor $k$ and bond dimension $\chi$ to a CFT at a finite temperature, where instead of minimizing the energy of the boundary state, one minimizes the free energy. In the CFT, turning on a temperature introduces a scale, going as the inverse temperature, which screens long-range correlations. Thus, the state will have classical correlations in addition to entanglement and the effect of a finite temperature on the entanglement entropy is the appearance of an extensive contribution. As one runs the MERA and coarse-grains, the thermal correlations that cannot be removed become more relevant. The MERA, which is unable to remove the extensive contribution, truncates at a level with multiple sites. The schematic entanglement renormalization process is illustrated in \Fig{fig:thermalMERA}. The state at the top level effectively factorizes, where each factor appears maximally mixed \cite{Swingle2012,Swingle2012a}. A tractable realization of this tensor network structure recently appeared in \Ref{Evenbly15}, which found a MERA representation of a thermal state.

Keeping in mind that the holographic dual of a finite-temperature state in the CFT is a black hole in AdS, where the temperature of the CFT corresponds to the Hawking temperature of the black hole, we note that the truncated MERA is suggestive of a black hole horizon \cite{Swingle2012}. If the MERA is to be interpreted as a discretization of the geometry, then the geometry has ended at some scale. Also, as we approach the horizon, the amount of Hawking radiation that we see increases and the temperature measured by an observer at the horizon diverges. The density matrix of some system in the infinite-temperature limit is given by the product of a maximally mixed state at each site, just like the state at the top of the MERA. It is important to note that, as was pointed out in \Ref{Evenbly15}, in order to reproduce the correct thermal spectrum of eigenvalues, a small amount of entanglement must be present between the sites at the horizon. If the bond dimension were taken to be infinite, then the sites at the horizon truly would factorize. But for a finite bond dimension, one should really think of the horizon as a high-temperature state, with sites effectively factorized.

For small regions on the boundary, the length of the subtending bulk geodesic is subextensive and so the Ryu--Takayanagi formula maintains that the boundary region's entanglement entropy is subextensive as well. However, if we consider a large enough region on the boundary, the geodesic will begin to probe the horizon of the black hole. The geodesic will run along the black hole horizon and pick up an extensive contribution to the entropy. We consider a boundary theory living on a lattice consisting of $n_{\rm b}$ sites, with total system coordinate length $x_{\rm sys} = n_{\rm b} a$.  In the limit as $r$ approaches the boundary in the metric \eqref{eq:BTZmetric}, we see that $Tx_{\rm sys} = r_+/L$, as was pointed out in Refs.~\cite{RyuTakayanagi, RyuTakayanagi2}. We further note that this implies that the system coordinate size is of order AdS radius, $x_{\rm sys} = 2\pi L$.

Let us now view the MERA of \Fig{fig:thermalMERA} as a discretization of a BTZ spacetime and repeat the analysis of \Sec{sec:geodesics}. In this discretization, the layers of the MERA lie along circles of fixed radius $r$ in the coordinates of \Eq{eq:BTZmetric}. Again, we ask what proper length $L_1$ separates sites in any given layer of the MERA.

First, note that a path at fixed $r_0$ that subtends an angle $\phi_0$ has proper length $r_0 \phi_0$. At the boundary of the MERA, we consider a region defined by $0\leq \phi \leq \phi_0 = 2\pi x_0/x_{\rm sys}$, where $x_0$ is the coordinate length of the interval, consisting of $\ell_0$ lattice sites. The boundary of the MERA is at a fixed radius $r=r_{\rm b}$. Naturally, the boundary radius $r_{\rm b}$ can be interpreted as a UV cutoff and is related to the lattice spacing $a$ by $r_{\rm b} = L^2/a$ \cite{RyuTakayanagi}. By equating the proper distance of the region in the MERA, $\ell_0 L_1$, with that at the boundary of the BTZ spacetime, $r_{\rm b} \phi_0$, we find the proper length between horizontal bonds to be $L_1=L$.

With the foresight that the top of the MERA is suggestive of a black hole horizon with proper length $2\pi r_+$, the number of sites at the final layer is therefore $n_{\rm h} = 2\pi r_+/L$. This further tells us that the MERA truncates after a finite number of layers $m$, given by
\be 
m = \log_k \left(\frac{n_{\rm b}}{n_{\rm h}}\right) = \log_k \frac{1}{2\pi T a}\,.\label{eq:mlayers}
\ee 
This coincides with the conclusion in Refs.\,\cite{Evenbly15,Molina14} that the MERA representation of a thermal state is obtained after $\mathcal{O}(\log_k (1/T))$ iterations of coarse-graining. 

Now consider a region $B$ on the boundary consisting of $\ell_0$ sites and for which the corresponding geodesic contains a segment running along the BTZ horizon. The subextensive contribution to the entropy in the MERA is exactly as before, in which we pick up at most $\ln \chi$ from each bond we cut with the causal cone of the region $B$. Furthermore, we will now pick up an extensive contribution from the horizon, where the number of horizon sites within the causal cone is $\ell_{\rm h}$ and each such site in the product state on the horizon contributes maximally to the entropy by an amount $\ln \chi$. Combining the contributions, we find 
\begin{equation}
S_{\rm MERA}(B) = 4 \eta_B f_B(k) \log_k \left(\frac{\ell_0}{\ell_{\rm h}}\right) \ln \chi + \ell_{\rm h}\ln \chi\,.
\label{eq:trunMERA}
\end{equation}

Recall that the entanglement entropy of a single interval $B$ of coordinate length $x_0$ in a CFT at finite temperature \cite{CardyCalabrese} is given, up to a non-universal constant, by
\begin{equation}
S_{\rm CFT} (B) = \frac{c}{3} \ln \left( \frac{1}{\pi a T} \sinh \pi x_0 T \right)\,,
\label{eq:EEtemp}
\end{equation}
where $x_0$ is much smaller than the total system size $x_{\rm sys}$. The standard field-theoretic derivation of the above entropy is done by computing the Euclidean path integral on an $n$-sheeted Riemann surface and analytically continuing to find the von~Neumann entropy. The same result can be derived by computing geodesic lengths on spatial slices of BTZ spacetimes and making use of the Ryu--Takayanagi formula.

When $T \rightarrow 0$ in \Eq{eq:EEtemp}, we recover the usual result \eqref{eq:CCent}. In the $T\rightarrow \infty$  limit, the von~Neumann entropy gives the usual thermal entropy as entanglement vanishes. Taking $T x_0 \gg 1$, the leading and subleading contributions to the entanglement entropy are
\begin{equation}
S_{\rm CFT} =  \frac{c}{3} \pi x_0 T + \frac{c}{3} \ln \frac{1}{2\pi a T} \,,\label{eq:finiteT-CFT-entropy}
\end{equation}
where the first term is the thermal entropy for the region $B$. 

Now let us consider a finite-temperature CFT that is dual to a BTZ black hole with horizon temperature $T=r_+/2\pi L^2$. In terms of geometric MERA parameters, we find that \Eq{eq:finiteT-CFT-entropy} becomes
\begin{equation}
S_{\rm CFT} =  \frac{c}{6} \, \ell_{\rm h} + \frac{c}{3} m \ln k  \, .\label{eq:finiteT-CFT-entropy-sub}
\end{equation}
Here we used the fact that $\ell_{\rm h} = x_0r_+/L^2$ as well as \Eq{eq:mlayers}, where we note that $m$ can also be written as $\log_k (\ell_{\rm b}/\ell_{\rm h})$. The result \eqref{eq:finiteT-CFT-entropy-sub} coincides precisely with the extensive and subextensive contributions calculated using the MERA in \Eq{eq:trunMERA} provided that $c/\ln \chi \sim \mathcal{O}(1)$. Therefore, we find that the truncated MERA correctly captures the entanglement structure of thermal CFT states and their dual BTZ spacetimes. 
These conclusions are in agreement with those in \Refs{MaldHart,Molina14}.

As a check of the claim that $c$ and $\ln\chi$ should be of the same order, we can compare the horizon entropy given by the contribution from the sites at the final layer with the Bekenstein--Hawking entropy \eqref{eq:BTZentropy} of a BTZ black hole. There are $n_{\rm h}$ sites comprising the horizon, each with Hilbert space dimension $\chi$. The system is in the infinite-temperature limit --- and hence described by a maximally mixed density matrix, with entropy contribution $\ln \chi$ from each site --- so
\be \label{eq:MERAhorS}
S_{\rm horizon} = n_{\rm h} \ln \chi\,.
\ee
Making use of the Brown--Henneaux relation and requiring that the entropy \eqref{eq:MERAhorS} coincide with the Beckenstein-Hawking entropy, we again find that $c/\ln \chi \sim \mathcal{O}(1)$. More specifically, taking the counting to be precise, we find that
\be c/\ln \chi = 6\,,\label{eq:chi-c-B}
\ee 
which is qualitatively in agreement with the previous conclusion \eqref{eq:chargeconstraint} that the Hilbert space dimension must be exponentially large in $c$.

With this relation, the extensive terms in \Eqs{eq:trunMERA}{eq:finiteT-CFT-entropy-sub} agree precisely. Further identifying the subextensive terms, we find $\eta_B f_B(k) = (\ln k)/2$. If we then impose the constraint \eqref{eq:eta-k}, we find that
\begin{equation}
\frac{k \ln k}{2(k-1)} \leq 1\, .
\end{equation}
This last inequality exactly reproduces Eq.~\eqref{eq:BoussoLoose} and thus constrains $k$ to be $2$, $3$, or $4$. Interestingly, we have found the weaker of the two bounds derived in \Sec{sec:BH}, without needing to consider the Bousso bound.

As desired, the truncated MERA computation of entanglement entropy agrees with the expected entanglement entropy given by the application of the Ryu--Takayanagi formula to the length of the minimal surface in a BTZ spacetime. The fact that the results of matching boundary entanglement entropy given in \Sec{sec:entanglement} further hold in BTZ spacetimes might not be too surprising given that such spacetimes are quotients of pure AdS$_3$.

\bibliographystyle{JHEP}
\bibliography{AdS-MERA}

\end{document}